\documentclass[aps,preprint,supberscriptaddress,showpacs,preprintnumbers,amsmath,amssymb,nofootinbib,longbibliography]{revtex4-1}
\usepackage[
colorlinks=true, 
pdfstartview=FitV, 
linkcolor=blue, 
citecolor=magenta, 
urlcolor=blue, 
bookmarks=true,
bookmarksnumbered=true,
pdftitle={},
pdfauthor={Masaru Hongo (RIKEN)}
]{hyperref}

\usepackage{amsmath,amssymb,bm}
\usepackage{graphicx}
\usepackage{bbold}
\usepackage{color}
\usepackage[normalem]{ulem}  
\usepackage[option]{colortbl}

\newcommand{\Tr}{\mathop{\mathrm{Tr}}}
\newcommand{\hH}{\hat{H}}
\newcommand{\hN}{\hat{N}}
\newcommand{\hK}{\hat{K}}
\newcommand{\hT}{\hat{T}}

\newcommand{\hrho}{\hat{\rho}}
\newcommand{\ptc}[1]{{\bar{#1}}}

\newcommand{\hS}{\hat{S}}
\newcommand{\hJ}{\hat{J}}

\newcommand{\hc}{\hat{c}}
\newcommand{\hcurrent}{\hat{\mathcal{J}}}

\newcommand{\hp}{\hat{p}}

\newcommand{\lie}{\mathsterling}

\newcommand{\hpi}{\hat{\pi}}

\newcommand{\hn}{\hat{n}}
\newcommand{\px}{\ptc{x}}
\newcommand{\pt}{\ptc{t}}
\newcommand{\timefunc}{\pt}

\newcommand{\dilatation}{\sigma}
\newcommand{\cA}{{\cal A}}

\newcommand{\covDer}{\nabla}

\newcommand{\hkappa}{\hat{\kappa}}

\newcommand\Lcal{\mathcal{L}}
\newcommand\Ocal{\mathcal{O}}
\newcommand\Dcal{\mathcal{D}}

\newcommand{\average}[1]{\langle#1\rangle}
\newcommand{\averageLG}[1]{\langle#1\rangle^\text{LG}}
\newcommand{\AverageLG}[1]{\Bigl\langle#1\Bigr\rangle^\text{LG}}

\newcommand{\la}[1]{\overleftarrow{#1}}
\newcommand{\ra}[1]{\overrightarrow{#1}}

\begin{document}

\preprint{RIKEN-STAMP-30}

\author{Masaru Hongo}

\email{masaru.hongo@riken.jp}

\affiliation{
iTHES Research Group, RIKEN, Wako 351-0198,
 Japan
}

\date{\today}
\title{Path-integral formula for local thermal equilibrium}

 \begin{abstract}
  We develop a complete path-integral formulation of 
  relativistic quantum fields in local thermal equilibrium, 
  which brings about the emergence of thermally induced curved spacetime.
  The resulting action is shown to have full diffeomorphism
  invariance and gauge invariance in thermal spacetime
  with imaginary-time independent backgrounds.
  This leads to the notable symmetry properties of emergent thermal spacetime: 
  Kaluza-Klein gauge symmetry, spatial diffeomorphism symmetry,
  and gauge symmetry. 
  A thermodynamic potential in local thermal equilibrium,
  or the so-called Masseiu-Planck functional, is identified as
  a generating functional for conserved currents
  such as the energy-momentum tensor and the electric current.
   
  \end{abstract}
\maketitle

\tableofcontents

\section{Introduction}
\begin{figure}[b]
 \centering
 \includegraphics[width=0.9\linewidth]{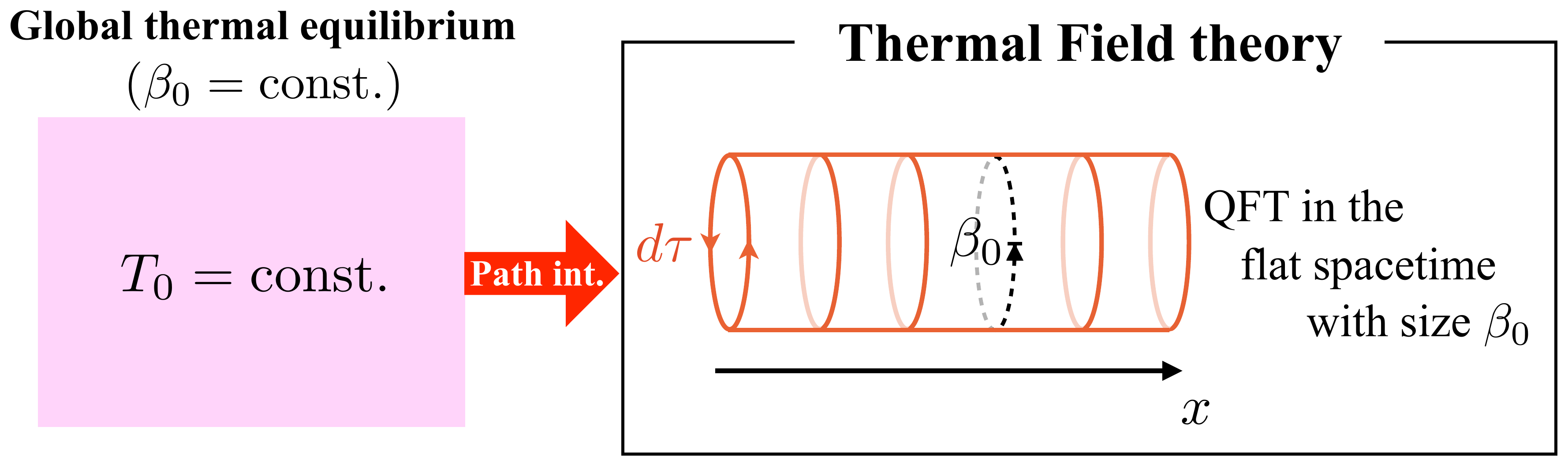}
 \caption{
 Schematic picture of the imaginary-time formalism.
} 
\label{Fig:imaginary-time}
\end{figure}

Finite temperature field theory is a tool indispensable for
theoretical study of many-body systems under global thermal
equilibrium~\cite{Kapusta,le2000thermal,KapustaGale}.
Its application to systems composed of relativistic quantum fields  
has uncovered rich aspects of the state of matter
in high energy physics such as 
the nature of the QCD phase transition from hadrons to the 
quark-gluon plasma (QGP), and the electroweak phase transition with its impact on
baryogenesis~(See
e.g. Refs.~\cite{
Aoki:2006we,Fukushima:2010bq,Ding:2015ona}
and references therein for reviews on the finite temperature QCD
and Refs.~\cite{Cohen:1993nk,Rubakov:1996vz,Morrissey:2012db}
on the electroweak phase transition). 
A standard formulation of the finite temperature field theory
is the imaginary-time formalism~\cite{Matsubara,AGD}, 
or the so-called Matsubara formalism. 
On the basis of the Gibbs ensemble,
it is formulated in terms of quantum field theories on a
compactified flat Euclidean spacetime, whose compact imaginary-time 
size is given by the inverse temperature
(See Fig.~\ref{Fig:imaginary-time}). 
It enables us to calculate the thermodynamic potential, 
in which all the information on the thermodynamic properties of systems
is fully contained. 
Once we evaluate the thermodynamic potential,
we can easily extract necessary information on thermodynamic properties
by taking its variation with respect to thermodynamic variables. 
However, it has an inherent limitation: namely, its application is,
by definition, restricted to systems in global thermal equilibrium. 

Although we have not yet had a general theoretical framework
applicable to nonequilibrium systems
beyond the linear response regime~\cite{Kubo2,Zwanzig},
there exists a well-established universal framework
if we restricts ourselves around local thermal equilibrium. 
In fact, in the vicinity of local thermal equilibrium, 
we can apply hydrodynamics to describe macroscopic behaviors
of the system~\cite{Landau:Fluid}, which captures the full nonlinear
spacetime evolution of the conserved charge densities. 
For example, the hydrodynamic description of the QGP created 
in heavy-ion collision experiments is one most
successful application of relativistic hydrodynamics~(See
e.g. Refs.~\cite{Kolb:2003dz,Romatschke:2009im,Teaney:2009qa,Gale:2013da}
and references therein
for reviews on the application of relativistic hydrodynamic to the QGP). 
However, compared to successful applications of hydrodynamics, 
its foundation from quantum field theories remains to be
understood.

For the purpose of understanding hydrodynamics based on
underlying microscopic theories, 
the first thing we have to do is to establish
a theoretical way to describe systems in local thermal equilibrium. 
The introduction of the \textit{local} Gibbs ensemble provides us a solid
basis to describe locally thermalized matter
as the (global) Gibbs ensemble does in the case of global equilibrium. 
As is discussed in Ref.~\cite{Hayata:2015lga} in the case of
the real scalar field,
the path-integral formula obtained with the local Gibbs ensemble method
has the same symmetry properties as the hydrostatic generating functional
method~\cite{Banerjee:2012iz,Jensen:2012jh}%
\footnote{
 See Refs.~\cite{Haehl:2015pja,Crossley:2015evo} for further recent developments for
 generalizations to incorporate non-hydrostatic and dissipative effects.  
}.
In spite of the same symmetry properties, 
physical situations under consideration differ between
the local Gibbs ensemble method and hydrostatic generating functional method.
Compared to the hydrostatic generating functional method,
the local Gibbs ensemble method has several advantages. 
In fact, the local Gibbs distribution 
is not restricted to the hydrostatic situations where
the local thermodynamic parameters only take stationary configurations.
Furthermore, it enables us to lay out a quantum field theoretical way to
calculate thermodynamic and transport properties of
locally thermalized matter based on underlying quantum field theories. 
In addition, we can elucidate the emergence of the thermally induced 
curved spacetime, and its relation to the local thermodynamic variables
without using matching condition to hydrodynamics.
All of these provides us a starting point
to derive hydrodynamic equations based on underlying
quantum field theories.

In this paper, we develop a complete path-integral formulation
of relativistic quantum fields in local thermal equilibrium
by the use of the local Gibbs distribution, 
which gives a robust extension of the imaginary-time formalism. 
In particular,
we derive the path-integral formula 
for a thermodynamic potential, or the so-called Masseiu-Planck functional, 
and show that it is regarded as the generating functional of 
locally thermalized systems. 
Our path-integral analysis shows
that the Masseiu-Planck functional is written 
in terms of quantum field theories 
in the emergent curved spacetime geometry, whose structure is completely
determined by configurations of the local thermodynamic variables. 
Furthermore, we demonstrate that, regardless of the spin of quantum fields, 
this emergent curved spacetime has the universal symmetry properties:
Kaluza-Klein gauge symmetry, spatial diffeomorphism symmetry, and gauge
symmetry.
These results provide a general microscopic justification and
generalization of the aforementioned
generating functional method~\cite{Banerjee:2012iz,Jensen:2012jh}
to the situations without hydrostatic conditions on the basis of nonequilibrium
statistical mechanics.

This paper is organized as follows: 
In Sec.~\ref{sec:LocalGibbs}, 
we review the local Gibbs distribution which describes 
locally thermalized systems based on the quantum field theory.  
In Sec.~\ref{sec:MasseiuPlanck}, we derive the variational formula for the
Massieu-Planck functional which enables us to extract information on
the average values of conserved current operators. 
In Sec.~\ref{sec:PathIntegral}, we provide explicit path-integral
formula for representative quantum fields such as
scalar fields, Dirac field, and gauge fields. 
In Sec.~\ref{sec:Symmetry}, we discuss the intrinsic symmetry arguments
of the Masseiu-Planck functional attached to the local Gibbs distribution. 
Section \ref{sec:Summary} is devoted to the summary and outlook.

\section{Preliminaries: Reviews on local Gibbs distribution}
\label{sec:LocalGibbs}
In this section we review our setup to describe locally thermalized
systems based on the quantum field theory. 
In Sec.~\ref{sec:ADM}, we first summarize the Arnowitt-Deser-Misner (ADM) 
decomposition of spacetime, which enables us to construct 
the local Gibbs distribution in a manifestly covariant way. 
In Sec.~\ref{sec:Matter}, we briefly review the relation between
symmetries and conservation laws of systems under external fields.  
In Sec.~\ref{sec:MaximalEntropy}, we introduce the local Gibbs 
distribution as the maximal entropy state from
the point of view of information theory. 

\subsection{Geometric preliminary}
\label{sec:ADM}
\begin{figure}[t]
 \centering
 \includegraphics[width=0.7\linewidth]{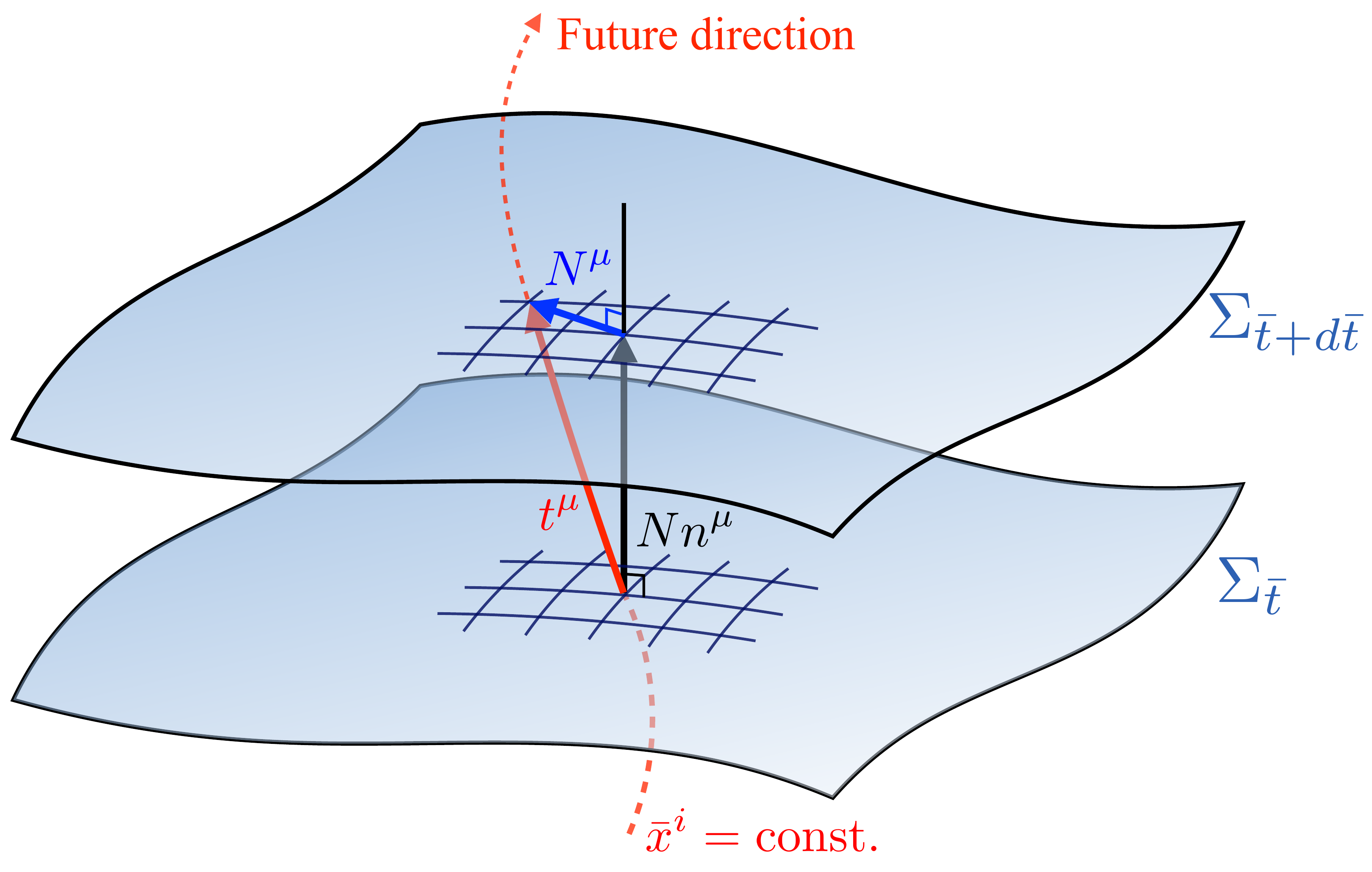}
 \caption{Illustration of the Arnowitt-Deser-Misner (ADM) decomposition of the  spacetime. $\Sigma_{\pt}$ denotes a spacelike hypersurface 
 parametrized by $\pt (x)=\ $const. $n^{\mu}$ is a vector normal to the hypersurface. 
 Introducing the lapse function $N(x)$ and the shift vector $N^{\mu}(x)$, 
 we decompose the time vector as $t^\mu \equiv \partial_\pt x^\mu  = Nn^\mu +N^\mu $.
 }
 \label{Fig:hypersurface}
\end{figure}

Let us consider a general curved spacetime, 
whose metric is given by $g_{\mu\nu}$%
\footnote{
As will be discussed in the subsequent section, the introduction
of the background metric helps us a lot
since it serves as an external source for the energy-momentum tensor.
}. 
We introduce the spatial slicings 
on this curved spacetime which are parametrized by a time coordinate 
function $\pt(x)$. 
We also introduce spatial coordinates $\bm{\px}= \bm{\px}(x)$ on these 
spacelike hypersurfaces. 
Here we have two important future-oriented timelike vectors: 
the unit normal vector $n_\mu$ perpendicular to the spacelike 
hypersurface, and the time vector $t^\mu$ 
which locally defines a time direction in our coordinate system. 
The ADM decomposition gives the decomposition of the time vector 
$t^\mu$ into the parts parallel and perpendicular to $n_\mu$
(See Fig.~\ref{Fig:hypersurface}). 
They are given as follows:
\begin{align}
 n_\mu(x) &= -N(x)\partial_\mu \timefunc(x)
 \quad \mathrm{with} \quad
 N(x)\equiv (-\partial^\mu \timefunc(x) \partial_\mu
 \timefunc(x))^{-1/2}\,. \\
 t^\mu (x) &\equiv \partial_\pt x^\mu (\pt,\bm{\px}) = N n^\mu + N^\mu 
 \quad \mathrm{with} \quad
  n_\mu N^\mu = 0.
\end{align}
Here the scalar function $N(x) > 0$ called the lapse function 
gives the normalization of the normal vector $n_\mu$ so as to satisfy 
$n_\mu n^\mu = -1$. 
We use the mostly plus convention of the metric,
e.g.,
the Minkowski metric is $\eta_{\mu\nu}\equiv\mathop{\mathrm{diag}} (-1,1,1,\cdots,1)$.
$N^\mu$ is the shift vector, which gives a perpendicular
part of the time vector to the normal vector. 
Here we note that $t^\mu$ can have the vortical distribution while
$n_\mu$ cannot from the Frobenius theorem. 
With the help of the normal vector, we define the hypersurface vector
$d\Sigma_{\pt\mu}$ as 
\begin{equation}
 d \Sigma_{\pt\mu} = - d\Sigma_\pt n_\mu 
  = - d^{d-1} \bm{\px} \sqrt{-g} N^{-1} n_\mu  ,
  \label{eq:HyperVector}
\end{equation}
where $d\Sigma_\pt = d^{d-1} \bm{\px}\sqrt{\gamma} $ denotes a volume element on the hypersuface 
with $\pt(x) = \mathrm{const.}$ 

In addition, we introduce the induced metric $\gamma_{\mu\nu}$ 
on the hypersurface as
\begin{align}
 \gamma_{\mu\nu} \equiv g_{\mu\nu}+n_\mu n_\nu. 
\end{align}
Then, if we express the metric $g_{\ptc{\mu}\ptc{\nu}}$ in 
our coordinate system $(\pt, \bm{\px})$, 
it becomes the form of the Arnowitt-Deser-Misner (ADM) metric,
\begin{equation}
 \begin{split}
  g_{\ptc{\mu}\ptc{\nu}} 
  = g_{\mu\nu} \frac{\partial x^\mu}{\partial \px^{\ptc{\mu}}}
  \frac{\partial x^\nu}{\partial \px^{\ptc{\nu}}}
  =
  \begin{pmatrix}
   -N^2 +N_\ptc{i}N^\ptc{i} & N_\ptc{j}\\
   N_{\ptc{i}}&  \gamma_{\ptc{i}\ptc{j}}
  \end{pmatrix},\quad
  g^{\ptc{\mu}\ptc{\nu}}=
  \begin{pmatrix}
   -N^{-2}& N^{-2}N^{\ptc{j}}\\
   N^{-2} N^{\ptc{i}}&  
   \gamma^{\ptc{i}\ptc{j}}-N^{-2}N^{\ptc{i}}N^{\ptc{j}}
  \end{pmatrix},
 \end{split}
 \label{eq:ADM}
\end{equation}
where $N_\ptc{i} =\gamma_{\ptc{i}\ptc{j}}N^\ptc{j}$.
Here $\gamma^{\ptc{i}\ptc{j}}$ denotes the inverse of 
$\gamma_{\ptc{i}\ptc{j}}$, and thus, satisfies 
$\gamma_{\ptc{i}\ptc{j}}\gamma^{\ptc{j}\ptc{k}}=\delta_\ptc{i}^\ptc{k}$.
By the use of the induced metric, 
we can express the $(d-1)$-dimensional volume element $d\Sigma_\pt$ on the hypersurface as 
\begin{equation}
 \int d\Sigma_\pt = \int d^d x \sqrt{-g} \delta(\pt - \pt(x)) N^{-1}(x) 
  = \int d^{d-1} \bm{\px} \sqrt{\gamma} , 
  \label{eq:VolumeElement}
\end{equation}
where we defined $\gamma = \det \gamma_{\ptc{i}\ptc{j}}$.
On the other hand, noting $\sqrt{-g} = N \sqrt{\gamma}$, 
the $d$-dimensional volume element is given by 
\begin{align}
 \int d^dx\sqrt{-g}=\int d^dxN\sqrt{\gamma}.
\end{align}

 \subsection{Matter field and conservation law}
\label{sec:Matter}
\subsubsection{Energy-momentum conservation law under external field}
Following the geometric preliminary, we next consider the matter sector and set out 
a general relation between symmetries of the system and conservation laws. 
We consider microscopic matter actions in a general curved spacetime background $g_{\mu\nu}$ and 
an external gauge field $A_\mu$, which is given by 
\begin{equation}
  S[\varphi; g_{\mu\nu}, A_\mu] = \int d^d x\sqrt{-g} \mathcal{L}(\varphi_i (x), \partial_\mu \varphi_i (x); g_{\mu\nu}(x), A_\mu(x)) ,
\end{equation} 
where $\varphi_i$ denotes a set of matter fields under consideration, 
and spacetime integral runs within all region in which matter fields
are placed%
\footnote{
If the action consists of spinor fields, it is not written in terms of the metric. 
Therefore, we have to take a slightly different way. 
Due to the extensive preparation required for that case, 
it will be discussed when we consider the Dirac field in Sec.~\ref{sec:Dirac}
}. 
Here we consider general situations with charged matter fields, 
but if matter fields are not charged, we do not have the external gauge field.

Since the action remains invariant under a general coordinate transformation, 
we have corresponding conserved charge currents associated with diffeomorphism invariance.
Let us consider the following infinitesimal coordinate transformation,
\begin{equation}
  x^\mu \rightarrow x'^\mu = x^\mu - \xi^\mu(x), 
  \label{eq:Diffeo}
\end{equation}
where $\xi^\mu(x)$ denotes an arbitrary infinitesimal vector. 
We assume that $\xi^\mu(x)$ vanishes on the boundary
region of spacetime integration for the action. 
Under the infinitesimal coordinate transformation \eqref{eq:Diffeo},
the variations of the metric $g_{\mu\nu}$, 
the external gauge field $A_\mu$, and matter fields $\varphi_i$
are given by Lie derivatives along $\xi^\mu$:
\begin{align}
  \lie_\xi g_{\mu\nu} &\equiv g_{\mu\nu}' (x) - g_{\mu\nu} (x)
  = \nabla_\mu \xi_\nu + \nabla_\nu \xi_\mu, \\
  \lie_\xi A_\mu &\equiv A_\mu' (x) - A_\mu (x) = \xi^\nu \nabla_\nu A_\mu 
  + A_\nu \nabla_\mu \xi^\nu, \\
  \lie_\xi \varphi_i &\equiv \varphi_i'(x) - \varphi_i (x) , 
\end{align}
where the explicit form of $\lie_\xi \varphi_i$ depends on the spin of fields such as 
$\lie_\xi \phi = \xi^\mu \partial_\mu \phi$ for the scalar field, 
and $\lie_\xi B_\mu = \xi^\nu \nabla_\nu B_\mu + B_\nu \nabla_\mu \xi^\mu$ for the vector field.
We, however, get rid of a change invoked by the variation of fields $\varphi_i$, 
with the help of the equation of motion for $\varphi_i$: $\delta S/\delta \varphi_i = 0$. 
Therefore, we obtain a following expression for the variation of the action,
\begin{equation}
  \begin{split}
   \delta S 
    &= \int d^d x \sqrt{-g} 
    \left[ \frac{1}{2} T^{\mu\nu}\lie_\xi g_{\mu\nu} 
      + J^\mu \lie_\xi A_\mu 
    \right] \\
    &= \int d^d x \sqrt{-g} 
    \left[ \frac{1}{2} T^{\mu\nu} (\nabla_\mu\xi_\nu + \nabla_\nu \xi_\mu)
      + J^\mu (\xi^\nu \nabla_\nu A_\mu + A_\nu\nabla_\mu \xi^\nu) 
    \right] \\
    &= -\int d^d x \sqrt{-g} 
    \left[ (\nabla_\mu T^{\mu}_{~\nu} - F_{\nu\lambda} J^\lambda  ) \xi^\nu
   \right] \\
   &\quad + \int d^d x \sqrt{-g}\nabla_\mu[(T^\mu_{~\nu} + J^\mu A_\nu)\xi^\nu]
    - \int d^d x \sqrt{-g} \xi^\nu A_\nu \nabla_\mu J^\mu ,
    \label{eq:deltaS}
  \end{split}
\end{equation}
where we defined the energy-momentum tensor $T^{\mu\nu}$ and charge current $J^\mu$ 
by taking variations of the action with respect to the metric and external gauge field:
\begin{align}
  T^{\mu\nu}(x) \equiv \frac{2}{\sqrt{-g}} \frac{\delta S}{\delta g_{\mu\nu}(x)}, \quad
  J^\mu(x) \equiv \frac{1}{\sqrt{-g}} \frac{\delta S}{\delta A_\mu (x)}.
  \label{eq:ConservedCurrent}
\end{align}
Here we also introduced a field strength tensor of the external gauge field $F_{\mu\nu}$ as 
\begin{equation}
  F_{\mu\nu} \equiv \partial_\mu A_\nu - \partial_\nu A_\mu.
\end{equation}
The second term in the last line of Eq.~\eqref{eq:deltaS} vanishes 
because it gives an integration on the boundary of the region,
where $\xi^\mu(x)$ does not take values.
Furthermore, the third term also vanishes due to the conservation law for the 
charge current as will be explained below.
Since the action is invariant ($\delta S =0$) under the above transformation with an arbitrary $\xi^\mu (x)$, 
Eq.~\eqref{eq:deltaS} results in the energy-momentum conservation law 
under the external field,
\begin{equation}
  \nabla_\mu T^\mu_{~\nu} = F_{\nu\lambda} J^\lambda.
  \label{eq:EMConservation}
\end{equation}
We note that this energy-momentum tensor is symmetric under $\mu \leftrightarrow \nu$ by definition%
\footnote{When we consider spinor fields, we use a vielbein $e_\mu^{~a}$ 
instead of the metric $g_{\mu\nu}$. 
We note that symmetry under $\mu \leftrightarrow \nu$ is not obvious in such a case. 
This is discussed in Sec.~\ref{sec:Dirac}. 
}.
This energy-momentum conservation law is an essential piece for
hydrodynamics.
We will use the operator version of this conservation law
in the subsequent discussion.

\subsubsection{Charge conservation law}
As is already mentioned, if systems contain charged matter fields, 
the charge current $J^\mu$ is also conserved. 
This stems from an internal symmetry of the action. 
Since this internal symmetry of the action is gauged, the action is
written in terms of the covariant derivative and
possesses gauge invariance under the $U(1)$ gauge
transformation%
\footnote{
In this paper, we only consider a single $U(1)$ symmetry
in the absence of quantum anomalies.
Generalizations to other conserved charges attached to non-Abelian gauge
symmetries are straightforward.
}
\begin{align}
 \delta_\alpha A_\mu &= \partial_\mu \alpha, \label{eq:GaugeTr1}\\
  \delta_\alpha \varphi_i &= i c_i \alpha \varphi_i, \label{eq:GaugeTr2}
\end{align}
where $\alpha(x)$ is an infinitesimal arbitrary function, which is assumed to be zero on the boundary, 
and $c_i = \pm 1,~0$ denotes the charge of matter fields $\varphi_i$.
The action is invariant under this $U(1)$ gauge transformation: $\delta S = 0$.
Then, let us express the variation of the action. 
Here we do not have to consider the variation of the matter fields,
because it does not contribute to the variation of the action by the use of 
the equation of motion: $\delta S/\delta \varphi_i = 0$.
Therefore, the variation of action is given by
\begin{equation}
  \begin{split}
    \delta S 
    & = \int d^d x \sqrt{-g} J^\mu \delta_\alpha A_\mu 
    = \int d^d x \sqrt{-g} J^\mu \partial_\mu \alpha \\
    & = -\int d^d x \sqrt{-g}\left[ (\nabla_\mu J^\mu) \alpha \right] 
    + \int d^d x \sqrt{-g}\nabla_\mu (J^\mu \alpha).
    \label{eq:deltaS2-1}
  \end{split}
\end{equation}
The second term in the second line of Eq.~\eqref{eq:deltaS2-1} is the boundary term, and again vanishes.
Since $\delta S=0$ holds for an arbitrary $\alpha(x)$,
we obtain the resulting conservation law
\begin{equation}
  \nabla_\mu J^\mu =0, 
  \label{eq:ChargeConservation}
\end{equation}
where $J^\mu$ is defined by the variation of the action with respect to the external gauge field in Eq.~\eqref{eq:ConservedCurrent}.

\subsection{Local Gibbs distribution as maximal information entropy state}
\label{sec:MaximalEntropy}
In this section, we introduce the local Gibbs distribution as 
the special form of the density operator
in which the information entropy functional 
takes the maximal value under the constraint on conserved charge densities~
\cite{Nakajima1,Nakajima2,Zubarev,Zubarev:1979,vanWeert1982133,Weldon:1982aq,Becattini:2014yxa}. 
Before discussing the local Gibbs distribution, we first consider 
the Gibbs distribution for global thermal equilibrium.

The vital point for equilibrium statistical mechanics is that 
we can express global thermal equilibrium by the use of 
the special state which does only depend on the value of conserved quantities 
such as the Hamiltonian $\hH$, and the conserved charge $\hN$. 
We can prepare such an appropriate expression of the density operator
by maximizing the information entropy (or von Neumann entropy):
\begin{equation}
 \begin{split}
  S(\hrho) =  -\Tr \hrho \log \hrho, 
 \end{split}
\end{equation}
under the following constraints
\begin{equation}
 \average{\hH} = E = \mathrm{const.}, \quad 
  \average{\hN} = N = \mathrm{const.}
\end{equation}
where the angle bracket denotes the average over $\hrho$: $\average{\hat{\Ocal}} \equiv \Tr \hrho \hat{\Ocal}$. 
In order to take into account the above constraints, 
we use a Lagrange multiplier method as usual. 
Then, our problem is to maximize the following quantity 
with respect to $\hrho$,
\begin{equation}
 S(\hrho;\lambda_E,\lambda_N) = - \Tr \hrho \log \hrho 
  + \lambda_E (\Tr \hrho \hH - E) + \lambda_N (\Tr \hrho \hN -N).
\end{equation}
where $\lambda_E$ and $\lambda_N$ are the Lagrange multipliers
conjugate to the energy $E$, and conserved charge $N$, respectively. 
Introducing the solution of the maximization problem as
$\hrho_{\mathrm{G}}$, we can write down the condition
for an extremal value as
\begin{equation}
 \Tr \delta \hrho
  \left( -\log \hrho_G + \lambda_E \hH + \lambda_N \hN  \right) = 0,
\end{equation}
where we used $\Tr \delta \hrho = 0$ coming from the normalization condition. 
This equation states that the solution takes the form of
$\hrho_G \propto e^{\lambda_E \hH + \lambda_N \hN}$. 
We, then, obtain the Gibbs distribution given by
\begin{equation}
  \hrho_G (\beta,\mu) = \frac{e^{-\beta (\hH  - \mu\hat{N})}}{Z} 
  \quad 
  \mathrm{with}
  \quad 
  Z= \Tr e^{-\beta(\hH -\mu \hat{N} )}, 
 \label{eq:GibbsDistribution1}
\end{equation}
where we used the familiar notation:
$\lambda_E = - \beta,~\lambda_N = \beta \mu $ with
the inverse temperature $\beta$ and the chemical potential $\mu$.
The partition function  $Z$ gives the normalization factor 
of the probability distribution, which is related to
the thermodynamic potential. 
By boosting the system, we can rewrite this distribution as
\begin{equation}
  \begin{split}
   \hrho_\mathrm{G} (\beta^\mu, \nu)
   = e^{\beta^{\mu}\hat{P}_\mu  +\nu \hat{N}-\Psi(\beta^\mu,\, \nu)},
   \quad \mathrm{with} \quad
   \Psi(\beta^\mu, \nu) \equiv
   \log \Tr e^{\beta^{\mu}\hat{P}_\mu  +\nu \hat{N} },
  \end{split}
  \label{eq:GibbsDistribution2}
\end{equation}
where parameters are $\beta^\mu=\beta u^\mu$ with
the global fluid four-velocity of the system $u^\mu$ normalized by
$u^\mu u_\mu=-1$, and $\nu = \beta \mu$. 
Here $\hat{P}_\mu$ denotes the total  energy-momentum of the system, and 
$\Psi(\beta^\mu,\nu)\equiv \log \Tr\exp[{\beta^{\mu}\hat{P}_\mu  +\nu \hat{N}}]$
is a thermodynamic potential\footnote{
According to Ref.~\cite{Kubo:Stat}, this thermodynamic function is called as the Kramers function. 
When we consider local thermal equilibrium in the subsequent discussion, 
we call the generalization of this thermodynamic function as the
Masseiu-Planck functional. 
}. 
Note that, at the rest frame of medium, $u^{\mu}=(1,\bm{0})$, and thus
$\hrho_\mathrm{G}(\beta^\mu, \nu) = \exp\bigl[{-\beta(\hat{H}  -\mu
\hat{N})-\Psi(\beta,\nu)}\bigr]$ is reproduced. 
This form of the density operator
is the so-called Gibbs distribution~\cite{Gibbs,Landau:Stat1}, and 
the corresponding ensemble is called the Gibbs ensemble, 
or the grand canonical ensemble.

We, then, generalize the global Gibbs distribution~\eqref{eq:GibbsDistribution2}
to local one in a manifestly covariant way~\cite{Zubarev:1979,vanWeert1982133,Weldon:1982aq,Becattini:2014yxa}. 
For this purpose, let us consider the density operator
at time $\pt$ which lives
on the spacelike hypersurface $\Sigma_\pt$. 
We can construct the local Gibbs distribution
by maximizing information entropy in a completely similar manner as before. 
In this case, to reproduce local thermodynamics on a given hypersurface
we have to put constraints on the average values of conserved charge 
densities on the hypersurface $\hc_a (x) \equiv \{\hp_\mu(x),~\hn'(x) \}$
--- the energy-momentum density
$\hp_\mu(x) \equiv - n_\nu(x) \hT^{\nu}_{~\mu}(x)$
and the conserved charge density $\hn'(x) \equiv -n_\nu (x) \hJ^\nu(x)$
--- as 
\begin{equation}
 \average{\hp_\mu(x)} =
  p_\mu(x) , \quad 
  \average{\hn'(x)} =  n'(x) .
\end{equation}
Here $n_\nu(x)$ is the normal vector, and
$\{\hT^\mu_{~\nu}(x),\,\hJ^\mu(x)\}$ a set of conserved current operators, 
defined in the previous section.
Note that while $\hp_\mu(x)$ and $\hn'(x)$ are the Heisenberg operators,
$p_\mu(x)$ and $n'(x)$ are $c$-number functions of the spatial coordinate on the hypersurface. 
Since we put the constraint on the average values of the local charge densities, 
the corresponding Lagrange multipliers become also local functions dependent on the spatial coordinate. 
Then, our problem is to maximize the following
quantity
\begin{equation}
 S[\hrho; \lambda_p^\mu, \lambda_n]
  \equiv -\Tr \hrho \log \hrho
  + \int d\Sigma_\pt 
  \Big[ \lambda_p^\mu(x) (\Tr \hrho \hp_\mu(x) - p_\mu(x)) 
  + \lambda_n(x) (\Tr \hrho \hn'(x) -  n'(x)) \Big] ,
\end{equation}
with respect to $\hrho$. 
Expressing the solution of the maximization problems as
$\hrho_{\mathrm{LG}}$, we obtain the following condition for an extremal value:
\begin{equation}
 \Tr \delta \hrho
  \left( -\log \hrho_{\mathrm{LG}}[\pt;\lambda]
   + \int d\Sigma_\pt  \lambda_p^\mu(x) \hp_\mu(x) 
   + \int d\Sigma_\pt \lambda_n(x) \hn'(x)  \right) = 0, 
\end{equation}
where we again used $\Tr \delta \hrho = 0$.
As a consequence, we obtain a local Gibbs distribution $\hrho_{\text{LG}}[\pt; \lambda]$ on the hypersurface as
\begin{equation}
    \hrho_{\text{LG}}[\pt; \lambda]\equiv\exp\bigl(-\hS[\pt;\lambda]\bigr)
    \quad
    \mathrm{with}
    \quad
    \hS[\pt;\lambda]\equiv\hK[\pt;\lambda]+\Psi[\pt;\lambda]\,,
  \label{eq:DefLG}
\end{equation} 
where $\hK[\pt;\lambda]$ is defined by
\begin{equation}
 \hK[\pt;\lambda] \equiv 
  -\int d\Sigma_{\pt \mu}\,\lambda^a(x)\hcurrent_a^\mu(x)
  =-\int d\Sigma_{\pt\nu}\, \Bigl( \beta^\mu(x){\hT^{\nu}}_{~{\mu}}(x)  +\nu(x)\hat{J}^{\nu}(x)\Bigr). 
  \label{eq:DefK}
\end{equation}
Here we used the hypersurface vector defined in Eq.~\eqref{eq:HyperVector}. 
We also defined a set of Lagrange multipliers $\lambda^a(x)\equiv\{\beta^\mu(x), \nu(x)\}$
with
$\lambda_p^\mu(x) \equiv \beta^\mu(x),~\lambda_n(x) \equiv \nu(x)$, and
a set of conserved current operators $\hcurrent_a^\mu(x)\equiv\{\hT^{\mu}_{~\nu}(x), \hJ^{\mu}(x)\}$, respectively.

As is the same as the global Gibbs distribution, a thermodynamic
potential $\Psi[\pt;\lambda]$, which is called the Massieu-Planck
functional, determines the normalization of the density operator 
$\hrho_{\text{LG}}[\pt;\lambda]$,
\begin{equation}
  \Psi[\pt;\lambda] \equiv \log \Tr \exp( - \hK[\pt;\lambda]).
  \label{eq:Masseiu-Planck}
\end{equation}
This is one of the most important quantities for the formulation of 
quantum field theories in local thermal equilibrium. 
It is a generalization of the thermodynamic function for systems under local thermal equilibrium, 
and gives fundamental variational formulae consistent
with local thermodynamics. 
In fact, if we take the variation of the Masseiu-Planck functional with respect to the 
local thermodynamic parameter $\lambda^a(x)$ on $\Sigma_\pt$, 
it gives the expectation values of the conserved charge densities 
$\averageLG{\hc_a (x)}_\pt\equiv \Tr\left[\hrho_{\text{LG}}[\pt; \lambda]\hc_a(x)\right]$:
\begin{equation}
  c_a(x)\equiv\averageLG{\hc_a(x)}_\pt=\frac{\delta}{\delta \lambda^a(x)}\Psi[\pt;\lambda]. \label{eq:Ca}
\end{equation}
Therefore, once the functional dependence of the Masseiu-Planck
functional on the thermodynamic parameters $\lambda^a(x)$ is known, 
it enables us to extract all the local thermodynamic properties of the
system like the speed of sound, the charge susceptibility,
and the equation of state. 
This is the reason why the Masseiu-Planck functional belongs to the family of 
the thermodynamic potentials
(See Ref.~\cite{Hayata:2015lga} for further discussions).
Furthermore, as is discussed in the next section,
it also provides useful variational formulae for the expectation values
of the conserved current operators over the local Gibbs distribution,
and thus, contains information on transport properties of
the locally thermalized matter.

In TABLE \ref{tab:Comparison}, we summarize our discussion
on local thermal equilibrium in comparison with global thermal equilibrium.  
We have shown that the local Gibbs ensemble method gives
a natural extension of the (global) Gibbs ensemble method, 
and they can be understood in a unified manner 
as a result of the maximization problem of the information entropy
functional under appropriate constraints.
\begin{table}[htb]
  \begin{tabular}{ccc} \hline \hline
    \rowcolor[rgb]{0.98,0.9,0.95}[2.015pt][2.015pt]
    \, \textbf{State/Statistical ensemble} \,  &
    \textbf{Characterized by} 	&
    \textbf{Conjugate variables}   \\ \hline  
    Global thermal equilibrium  &  
    Conserved charges  
    & Thermodynamics parameters    \\ 
    Gibbs ensemble \eqref{eq:GibbsDistribution2}
    & $C_a = \{ H, N  \}$ 
    &  $\lambda^a = \{\beta, \nu  \}$    \\ \hline
    Local thermal equilibrium 
    & Conserved charge densities 
    & \, Local thermodynamic parameters \,   \\ 
    Local Gibbs ensemble \eqref{eq:DefLG}
    & $c_a(x) = \{p_\mu(x), n' (x)  \}$ 
    & $\lambda^a (x) = \{\beta^\mu(x), \nu(x)  \}$   \\
    \hline \hline
  \end{tabular}
  \caption{Comparison of ways to describe systems under global/local thermal equilibrium 
    based on the Gibbs and local Gibbs ensembles.}
  \label{tab:Comparison}
\end{table}

\section{Masseiu-Planck functional as generating functional}
\label{sec:MasseiuPlanck}
In this section we derive a useful variational formula
which relates the Masseiu-Planck functional 
to the average value of the conserved current operators
$\hcurrent_a^\mu(x) = \{ \hT^{\mu\nu}(x),~ \hJ^\mu(x) \}$
over the local Gibbs distribution.
In Sec.~\ref{sec:VariationMP}, we provide the general derivation without
gauge fixing, or without choosing the specific coordinate system. 
In Sec.~\ref{sec:Hydrostatic}, we discuss the useful gauge choice,
which we call \textit{hydrostatic gauge}%
\footnote{
The same name for the similar situation is also employed in
Ref.~\cite{Haehl:2015pja}.},
and re-express the variational formulae in the hydrostatic gauge.

\subsection{Derivation of variational formula without gauge fixing}
\label{sec:VariationMP}
In order to derive the variational formula for the Masseiu-Planck functional,
we first turn our attention to the 
the following expression of
$\hK[\pt,\lambda^a,g_{\mu\nu},A_\mu]$: 
\begin{equation}
 \hK [\pt,\lambda^a, g_{\mu\nu},A_\mu]
  = \int d^d x \sqrt{-g} \delta(\pt - \pt(x) )
  \hkappa (\lambda^a,g_{\mu\nu},A_\mu),
  \label{eq:Khat1}
\end{equation}
where we used Eq.~\eqref{eq:VolumeElement} and 
defined 
\begin{equation}
  \hkappa (\lambda^a,g_{\mu\nu},A_\mu) 
  \equiv N^{-1}  n_\mu \lambda^a(x)\hcurrent_a^\mu(x) 
  = - N^{-1}(x) \hc_a (x) \lambda^a(x). 
\end{equation}
Since the spacetime integral in Eq.~\eqref{eq:Khat1} runs within
all region where matter fields exist,
$x$ is nothing but a dummy variable.
We, therefore, have reparametrization invariance of $x$. 
Let us consider the leading-order contribution
caused by an infinitesimal reparametrization:
$x^\mu \to x'^\mu = x^\mu - \xi^\mu(x)$. 
Since this reparametrization can be regarded as a coordinate transformation, 
invariance of $\hK[\pt,\lambda^a,g_{\mu\nu},A_\mu]$ brings about
\begin{equation}
 \hK [\pt + \lie_\xi \pt,\lambda^a + \lie_\xi \lambda^a ,
  g_{\mu\nu}+ \lie_\xi g_{\mu\nu}, A_\mu + \lie_\xi A_\mu]
  = \hK [\pt, \lambda^a ,g_{\mu\nu}, A_\mu]   ,
  \label{eq:identity}
\end{equation}
where $\lie_\xi$ denotes the usual Lie derivative along the arbitrary
vector $\xi^\mu(x)$ introduced in Sec.~\ref{sec:Matter}. 
This identity follows from
the transformation laws for $\pt$, $\lambda^a$, $g_{\mu\nu}$, and $A_\mu$:
\begin{align}
 \pt\, ' (x')
 &= \pt(x') + \lie_\xi \pt(x'), \\
 {\lambda^a}'(x')  &=
  \begin{cases}
   \beta^\mu(x') + \lie_\xi  \beta^\mu (x')
   \quad &(a=0,1,\cdots,d-1)
   \\
    \nu (x') + \lie_\xi \nu (x')  \quad &(a=d) 
\end{cases}, \\
 g'_{\mu\nu} (x')
 &= g_{\mu\nu}(x') + \lie_\xi g_{\mu\nu} (x'),  \\
 A_\mu' (x') &= A_\mu (x') + \lie_\xi A_\mu(x') .
\end{align}
By choosing $\xi^\mu(x) = \beta^\mu(x)$ in the above identity, 
and using gauge invariance, we have 
\begin{equation}
 \delta_\lambda \hK \equiv
  \hK [\pt + \lie_\beta \pt,\lambda^a + \lie_\beta \lambda^a ,
  g_{\mu\nu}+ \lie_\beta g_{\mu\nu}, A_\mu + \delta_\beta A_\mu]
  -  \hK [\pt ,\lambda^a ,g_{\mu\nu}, A_\mu] = 0,
  \label{eq:identity1}
\end{equation}
where we defined
$\delta_\beta A_\mu \equiv \lie_\beta A_\mu + \nabla_\mu \alpha$
with $\alpha \equiv \nu - \beta\cdot A$.
The second term in $\delta_\beta A_\mu$ is coming from
the gauge transformation: $A_\mu \to A_\mu + \nabla_\mu \alpha$. 
Then, we can express $\delta_\lambda \hK$ in terms of 
the variation with respect to the background metric
and gauge field as
\begin{equation}
 \begin{split}
  \delta_\lambda \hK
  &=  \int d^d x
  \left[ \frac{\delta \hK}{\delta \pt(x)} \lie_\beta \pt(x)
  + \frac{\delta \hK}{\delta \lambda^a(x)} \lie_\beta \lambda^a(x)
  + \frac{\delta \hK}{\delta g_{\mu\nu}(x)} \lie_\beta g_{\mu\nu}(x)
  + \frac{\delta \hK}{\delta A_\mu(x)} \delta_\beta A_\mu(x) \right]  \\
  &= \int d^d x
  \Bigg[
  \sqrt{-g} \left( \hT^\mu_{~\nu} (\nabla_\mu \beta^\nu)
  + \hJ^\mu (\nabla_\mu  \nu + F_{\nu\mu} \beta^\nu) \right)
  \beta^\mu \partial_\mu \pt
  \\ &\hspace{52pt}
  + \frac{\delta \hK}{\delta g_{\mu\nu}} \lie_\beta g_{\mu\nu}
  + \left( \frac{\delta \hK}{\delta \nu} \beta^\mu 
  + \frac{\delta \hK}{\delta A_\mu} \right) \delta_\beta A_\mu \Bigg]
  \\
  &= \int d^d x
  \left[
  \left( \frac{\beta^\ptc{0}\sqrt{-g}}{2} \hT^{\mu\nu}
  + \frac{\delta \hK}{\delta g_{\mu\nu}} \right) \lie_\beta g_{\mu\nu}
  + \left( \beta^\ptc{0} \sqrt{-g} \hJ^\mu
  + \frac{\delta \hK}{\delta \nu} \beta^\mu 
  + \frac{\delta \hK}{\delta A_\mu}
  \right) \delta_\beta A_\mu \Big]
  \right],
 \end{split}
\end{equation}
where we employed another expression of
$\hK[\pt,\lambda^a,g_{\mu\nu},A_\mu]$
\begin{equation}
 \hK [\pt,\lambda^a, g_{\mu\nu},A_\mu]
  = \int d^d x \sqrt{-g} \theta(\pt - \pt(x))
  \nabla_\mu ( \lambda^a(x)\hcurrent_a^\mu(x) ) ,
  \label{eq:Khat}
\end{equation}
and used the conservation laws for current operators
\eqref{eq:EMConservation} and \eqref{eq:ChargeConservation}
to rewrite the first term in the first line. 
We also used a set of relations like $\lie_\beta \beta^\mu = 0$, 
$\lie_\beta \nu = \lie_\beta \alpha + \beta^\mu \lie_\beta A_\mu
= \beta^\mu \delta_\beta A_\mu$, and
$\nabla_\mu  \nu + F_{\nu\mu} \beta^\nu = \delta_\beta A_\mu$. 
Taking average of $\delta_\lambda\hK$ over the local Gibbs distribution
at that time, and replacing the average values of variations of $\hK$
with the variations of the  Masseiu-Planck functional,
we obtain 
\begin{equation}
 \begin{split}
  \averageLG{\delta_\lambda \hK}_\pt
  &= \int d^d x
  \Bigg[
  \left( \frac{\beta^\ptc{0}\sqrt{-g}}{2} \averageLG{\hT^{\mu\nu}}_\pt
   + \frac{\delta \Psi}{\delta g_{\mu\nu}} \right) \lie_\beta g_{\mu\nu}
  \\ & \hspace{52pt}
  + \left( \beta^\ptc{0} \sqrt{-g} \averageLG{\hJ^\mu}_\pt
  + \frac{\delta \Psi}{\delta \nu} \beta^\mu 
  + \frac{\delta \Psi}{\delta A_\mu}
  \right) \delta_\beta A_\mu \Big]
  \Bigg]. 
 \end{split}
\end{equation}
The identity \eqref{eq:identity1} results in
$\averageLG{\delta_\lambda \hK}_\pt = 0$ for an arbitrary variation
of the background metric and gauge field, and thus,  
it immediately enables us to relate the average values of
the conserved current operators 
over the local Gibbs distribution 
with the variation of the Masseiu-Planck functional:
\begin{align}
  \averageLG{\hT^{\mu\nu}(x)}_\pt 
  = \frac{2}{\beta^\ptc{0}\sqrt{-g}} \frac{\delta\Psi[\pt;\lambda]}{\delta
 g_{\mu\nu}(x)}  ,\quad 
  \averageLG{\hJ^{\mu}(x)}_\pt 
  = \frac{1}{\beta^\ptc{0}\sqrt{-g}} 
 \left( \frac{\delta \Psi[\pt;\lambda]}{\delta \nu(x)} \beta^\mu(x)
 + \frac{\delta \Psi[\pt;\lambda]}{\delta A_{\mu}(x)}  \right).
 \label{eq:VariationFormula1}
\end{align}
Therefore, the expectation values of all kinds of conserved current operators
in local thermal equilibrium is captured by the
single functional, or the Masseiu-Planck functional.  
Since we introduce the background field, the above variational formula
seemingly looks ordinary one.
However, compared to Eq.~\eqref{eq:ConservedCurrent} which is nothing but 
the definition of the conserved current operators,
Eq.~\eqref{eq:VariationFormula1} does not provide the definition
but the relation between the expectation values of the conserved current operators
and the Masseiu-Planck functional. 
This relation is not obvious at all,
as is demonstrated in the first term of
$\averageLG{\hJ^\mu(x)}_\pt$ in Eq.~\eqref{eq:VariationFormula1}.
In conclusion, we can identify the Masseiu-Planck functional
as a generating functional for nondissipative hydrodynamics,
in which we neglect the deviation from the local Gibbs distribution at each time.

\subsection{Useful gauge choice: Hydrostatic gauge}
\label{sec:Hydrostatic}
In the previous section, we derive the variational formula 
without a specific choice of the coordinate system 
by considering reparametrization invariance of
$\hK[\pt,\lambda^a,g_{\mu\nu},A_\mu]$. 
Here, by considering the derivative of $\Psi[\pt,\lambda]$
with respect to $\pt$, and choosing our coordinate system
so that the time vector $t^\mu \equiv \partial_\pt x^\mu(\pt,\bm{\px}) $ is along
the fluid vector $\beta^\mu$,
we rederive the variational formula. 
First of all, the time derivative of $\Psi[\pt,\lambda]$ reads
\begin{equation}
  \begin{split}
   \partial_\pt \Psi[\pt;\lambda]
   &= -\averageLG{\partial_\pt \hK[\pt;\lambda] }_{\pt} 
   =  \AverageLG{\partial_\pt\int d\Sigma_{\pt\mu}
   \lambda^a\hcurrent_a^\mu}_\pt\\
   &= \AverageLG{\int d\Sigma_\pt N
   \covDer_\mu \bigl( \lambda^a\hcurrent_a^\mu\bigr)}_\pt \\
   &= \int d^{d-1} \bm{\px} \sqrt{-g} 
   \left(  \nabla_\mu \beta_\nu  \averageLG{\hT^{\mu\nu}}_\pt
   +(\nabla_\mu\nu + F_{\nu\mu}\beta^\nu)\averageLG{\hJ^\mu}_\pt \right) 
   \label{eq:delPsi}
  \end{split}
\end{equation} 
where we used the conservation laws
in Eqs.~\eqref{eq:EMConservation} and \eqref{eq:ChargeConservation}. 
We also used 
\begin{equation}
    \partial_\pt \int d\Sigma_{\pt\mu}  f^\mu =  \int d\Sigma_\pt N \covDer_\mu f^\mu,  \label{eq:StokesTheorem}
\end{equation}
which holds for an arbitrary smooth function $f^{\mu}(x)$%
\footnote{
As an alternative, we can obtain the same result by the use
of the expression of $\hK[\pt,\lambda]$ in Eq.~\eqref{eq:Khat}, 
}.

\begin{figure}[t]
 \centering
 \includegraphics[width=1.0\linewidth]{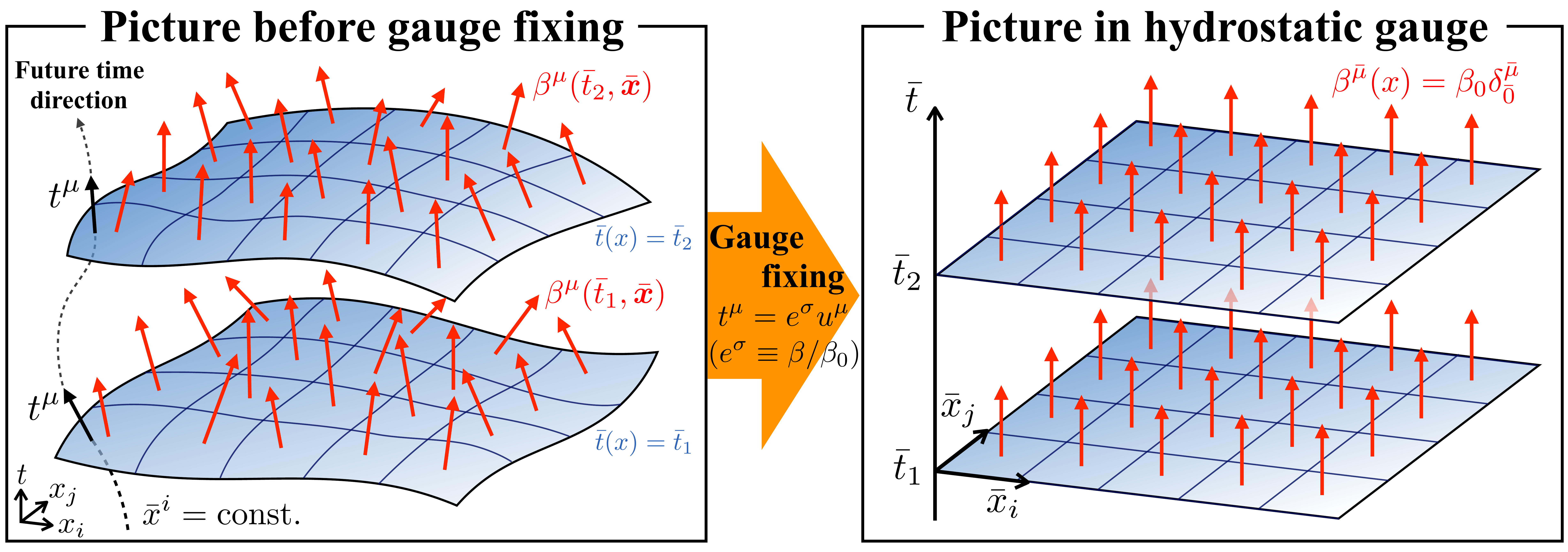}
 \caption{
  Schematic picture of a gauge fixing to the hydrostatic gauge. 
  Before the gauge fixing, we have a fluid configuration $\beta^\mu(x)$ on the hypersurface, 
  which could be any time-like vector. 
  After the gauge fixing by choosing $t^\mu (x)= \beta^\mu (x)$, it simply becomes a set of the unit vectors. 
} 
\label{Fig:hydrostatic-gauge}
\end{figure}

To take one more step forward, we choose the useful coordinate system 
by matching the time vector $t^\mu (x)\equiv \partial_\pt x^\mu(\pt,\bm{\px})$ with the local fluid vector 
$\beta^\mu (x)$: $t^\mu(x) = \beta^\mu(x)/ \beta_0 $,
where $\beta_0$ is some constant reference temperature. 
We can arbitrary choose the value of $\beta_0$, 
among which the most useful choice is to adopt the value 
when the system reach global thermal equilibrium.
Besides, we interpret the chemical potential as the time
component of the background $U(1)$ gauge field: 
$\nu (x) = A_\nu(x) t^\nu (x) = A_\ptc{0}(x)$. 
Then, the hydrostatic condition can be summarized as follows: 
\begin{equation}
 t^\mu(x) = \beta^\mu(x)/ \beta_0 , \quad
  A_\ptc{0}(x) = \nu(x).
  \label{eq:HydroStaticCond}
\end{equation}
This gauge fixing is schematically shown in Fig.~\ref{Fig:hydrostatic-gauge}. 
Before adopting this gauge choice, spacetime coordinates have nothing to
do with hydrodynamic configurations $\lambda^a(x)$.
After gauge fixing, the fluid vector $\beta^{\ptc{\mu}}$ becomes 
a future-time directed constant vector in the new coordinate system, 
and spacetime coordinates are fully related to hydrodynamic
configurations through
the hydrostatic gauge condition \eqref{eq:HydroStaticCond}%
\footnote{
Indeed, the hydrostatic gauge condition $t^\mu(x) \equiv
\partial_\pt x^\mu (\pt,\bm{\px})=
\beta^\mu(x)/ \beta_0 $ indicates that the original coordinate
$x^\mu(\pt,\bm{\px})$ in the hydrostatic gauge plays a role as the label of
fluid parcels (particles) in the Lagrangian specification
(See e.g. Ref.~\cite{Soper} for a review on the Lagrangian description
of the relativistic fluid). 
}.  
Since the fluid remains at rest in this coordinate system, we call it 
the hydrostatic gauge%
\footnote{
 Our hydrostatic space (See the right picture in
 Fig.~\ref{Fig:hydrostatic-gauge}) corresponds to the reference frame
 in Ref.~\cite{Haehl:2015pja}, and fluid spacetime in Ref.~\cite{Crossley:2015evo}
}. 
Under this parametrization, we obtain 
\begin{equation}
 \begin{split}
  \partial_\pt \Psi [\pt;\lambda]
  &= \int d^{d-1} \bm{\px} \sqrt{-g} 
  \left( \frac{1}{2}  \averageLG{\hT^{\mu\nu}}_\pt
  (\nabla_\mu \beta_\nu + \nabla_\nu \beta_\mu )
  +  \averageLG{\hJ^\mu}_\pt
  ( \beta^\nu \nabla_\nu A_\mu + A_\nu \nabla_\mu \beta^\nu   )
  \right) \\
  &= \int d^{d-1} \bm{\px} \sqrt{-g} 
  \left( \frac{1}{2}  \averageLG{\hT^{\mu\nu}}_\pt \lie_\beta g_{\mu\nu} 
  +  \averageLG{\hJ^\mu}_\pt \lie_\beta A_\mu \right),  
   \label{eq:VariationMPF1}
 \end{split}
\end{equation}
where we used the symmetry of energy-momentum tensor under
$\mu \leftrightarrow \nu$. 
Here $\lie_\beta$ again denotes the Lie derivatives
along the fluid-vector $\beta^\mu(x)$. 

On the other hand, the variation of the Massieu-Planck functional
with respect to $\pt$, or the Lie derivative along $t^\mu(x)$,
is expressed in another way.
Since we choose $t^\mu(x) = \beta^\mu(x)/\beta_0$ in the hydrostatic gauge,
we can use $\lie_t \beta^\mu = (\lie_\beta \beta^\mu)/ \beta_0 = 0$.
In addition, the second condition in Eq.~\eqref{eq:HydroStaticCond},
or $\nu(x)= A_\ptc{0}(x)$, 
enables us to put the variation of $\nu(x)$ and $A_\mu(x)$ all together.
As a result, we obtain another expression of $\partial_\pt \Psi[\pt;\lambda]$:
\begin{equation}
  \begin{split}
    \partial_\pt \Psi [\pt;\lambda] 
   &= \int d^{d-1} \bm{\px} \frac{1}{\beta_0}
   \left(  \frac{\delta \Psi}{\delta g_{\mu\nu}} \lie_\beta g_{\mu\nu}
      +\frac{\delta \Psi}{\delta A_\mu} \lie_\beta A_\mu 
    \right).
  \end{split}
  \label{eq:VariationMPF2}
\end{equation}
Comparison of Eq.~\eqref{eq:VariationMPF1} with Eq.~\eqref{eq:VariationMPF2} 
brings about the variational formula in the hydrostatic gauge as 
\begin{equation}
  \averageLG{\hT^{\mu\nu}(x)}_\pt  
  = \frac{2}{\beta_0 \sqrt{-g}}
   \frac{\delta \Psi[\pt;\lambda]}{\delta g_{\mu\nu}(x)}
  \Bigg|_{\mathrm{hs}}, \quad 
  \averageLG{\hJ^{\mu}(x)}_\pt
  = \frac{1}{\beta_0 \sqrt{-g}}
   \frac{\delta \Psi[\pt;\lambda]}{\delta A_{\mu}(x)} \Bigg|_{\mathrm{hs}}. 
 \label{eq:VariationFormula2} 
\end{equation}
We do not have the variation of $\Psi[\pt;\lambda]$ with
respect to $\nu(x)$ this time because one of the hydrostatic gauge condition
$\nu(x) = A_\ptc{0}(x)$ put it together with the variation
with respect to $A_\mu(x)$. 
Note that $\beta_0$ is not the time component of the fluid vector. 

Before closing this section, we put some comments on the hydrostatic gauge.  
First, although hydrodynamic configurations looks like entirely 
at rest in the hydrostatic gauge (See the right picture
in Fig.~\ref{Fig:hydrostatic-gauge}), 
it does not imply that our system is stationary at all. 
In fact, if the time vector in the hydrostatic gauge is not a killing vector:
$\lie_t g_{\mu\nu} =
( \nabla_\mu \beta_\nu + \nabla_\nu \beta_\mu)/\beta_0 \neq 0$, 
our system will evolve in time accompanied with a finite entropy production,
governed by hydrodynamic equations. 
Therefore, the system itself is not hydrostatic at all.
We also note that although we can always take the
hydrostatic gauge at any hypersurface, 
it does not mean that our density operator always has a exact form of the
local Gibbs distribution. 
In fact, as is discussed in Ref.~\cite{Hayata:2015lga},
the deviation from the local Gibbs distribution gives
proper dissipative corrections to hydrodynamic constitutive relations.
Another comment is about the spatial slicings.
Since the spatial hypersurface is characterized by the normal vector $n_\mu$,
it cannot have the vortical configuration due to the Frobenius theorem.
Therefore, we cannot always match the normal vector $n^\mu$ with
the fluid vector $\beta^\mu$ as is the case for the time vector $t^\mu$.
When we take hydrostatic gauge in such a case,
$t^\mu$ is not proportional to $n^\mu$, which means that
we cannot remove the shift vector $N^\mu$. 
The final comment on the hydrostatic gauge is related to
the subject given in the next section.
If we take the hydrostatic gauge, 
our background metric $g_{\mu\nu}$ and gauge field $A_\mu$
are related to the local thermodynamic parameters $\lambda^a(x)$, 
and, as a result, they coincide with the thermal metric 
$\tilde{g}_{\mu\nu}$, and gauge field $\tilde{A}_\mu$,
which will be introduced
in Eqs.~\eqref{eq:thermalMetric} and \eqref{eq:tildeA}.

\section{Path integral formula and emergent curved spacetime}
\label{sec:PathIntegral}
In this section, dealing with some representative examples
of quantum fields such as the scalar field, Dirac field, and gauge field,
we explicitly perform path-integral analysis
for the Masseiu-Planck functional $\Psi[\pt;\lambda]$. 
As a consequence, we show that the Masseiu-Planck functional is written in
terms of the Euclidean action in the same way as
the case of global thermal equilibrium%
\footnote{
To say it properly, it may not be appropriate to call it the Euclidean action 
since the action can be imaginary in the case of local thermal equilibrium. 
}.
It, however, does not have the form of that in the flat spacetime, 
but in the thermally emergent curved spacetime background, 
whose metric or vielbein and gauge field are determined
by the local temperature, and the fluid four-velocity. 
Furthremore, if matter fields under consideration is electrically charged,
there exists the background $U(1)$ gauge connection
which is determined by the local chemical potential.
Therefore, path-integral formula for the Masseiu-Planck functional is
given as follows:
\begin{equation}
 \Psi [\pt;\lambda] = \log \int \Dcal \varphi \,
  e^{S[\varphi;\lambda]} \quad \mathrm{with} \quad
  S[\varphi;\lambda]
  = \int_0^{\beta_0}  d^d \tilde{x} \sqrt{-\tilde{g}}
  \tilde{\Lcal}
  (\varphi_i, \tilde{D}_\ptc{\rho}\varphi_i;\,
  \tilde{g}_{\ptc{\mu}\ptc{\nu}},\tilde{e}_\ptc{\mu}^{~a},\tilde{A}_\ptc{\mu}),
  \label{eq:PathIntegralFormula}
\end{equation}
where we defined
$\displaystyle{\int_0^{\beta_0} d^d \tilde{x} \equiv 
\int_0^{\beta_0} d\tau \int d^{d-1}\px}$ with
an arbitrary constant reference temperature $\beta_0$. 
Here $\tilde{g}_{\ptc{\mu}\ptc{\nu}}$, $\tilde{e}_\ptc{\mu}^{~a}$ and
$\tilde{A}_\ptc\mu$ denote the thermal metric, vielbein, and
external gauge field in the emergent thermal spacetime,
which are defined in Eqs.~\eqref{eq:thermalMetric}, \eqref{eq:tildee}
and \eqref{eq:tildeA}, respectively. 
We also introduced the proper covariant derivative
$\tilde{D}_\ptc{\rho}$ in the thermal spacetime, whose explicit form
are given in Eqs.~\eqref{eq:CovDer1}-\eqref{eq:CovDer2}
for the charged scalar field, and Eq.~\eqref{eq:CovDerFermion2}
for the Dirac field.
The vital point here is that the structure of the emergent thermal spacetime
and gauge connection is universal regardless of the spin of the
microscopic quantum fields.
We will show the explicit derivation of this path-integral formula
for each quantum field as follows.

\subsection{Scalar field}
\subsubsection{Real scalar field}
Let us first review the path-integral formula for a one-component
real scalar field derived in Ref.~\cite{Hayata:2015lga}. 
In the coordinate system $(\pt,\bm{\px})$ with the ADM metric~\eqref{eq:ADM}, the Lagrangian for a neutral scalar field $\phi$ reads
\begin{equation}
    \mathcal{L} = -\frac{g^{\ptc{\mu}\ptc{\nu}}}{2}\partial_\ptc{\mu}\phi \partial_\ptc{\nu}\phi  - V(\phi)
    = \frac{1}{2N^2}(\partial_\pt\phi -N^{\ptc{i}}\partial_\ptc{i}\phi)^2-
    \frac{\gamma^{\ptc{i}\ptc{j}}}{2}\partial_{\ptc{i}}\phi\partial_{\ptc{j}}\phi-V(\phi) ,
\end{equation}
where $V(\phi)$ denotes the potential term.
The canonical momentum $\pi(\bm{x})$ is $\pi(\bm{x}) \equiv 
\partial (\sqrt{-g} \Lcal) /\partial (\partial_\ptc{0} \phi ) =   -
\sqrt{-g} g^{\ptc{0}\ptc{\nu}}\partial_\ptc{\nu}\phi(\bm{\px})= 
N^{-1}\sqrt{\gamma} (\partial_{\ptc{t}}\phi-N^{\ptc{i}}
\partial_{\ptc{i}}\phi)$, which satisfies the canonical commutation relation,
$[\hat{\phi}(\pt,\bm{\px}),\hpi(\pt,\bm{\px}')] =
i\delta(\bm{\px}-\bm{\px}')$%
\footnote{
We slightly changed the definition of the canonical momentum
from Ref.~\cite{Hayata:2015lga} to include $\sqrt{-g}$. 
}. 
As discussed in Sec.~\ref{sec:Matter}, we obtain
the energy-momentum densities as
\begin{align}
  {\hT^{\ptc{0}}}_{~\ptc{0}} 
  &= - \frac{1}{\sqrt{-g}}\hpi \partial_{\pt}\hat{\phi}+\hat{\mathcal{L}}
 = - \frac{1}{2\gamma} \hpi^2
 - \frac{N^{\ptc{i}}}{N\sqrt{\gamma}} \hpi \partial_{\ptc{i}}\hat{\phi}
  -\frac{\gamma^{\ptc{i}\ptc{j}}}{2}\partial_{\ptc{i}}\hat{\phi}\partial_{\ptc{j}}\hat{\phi}
  -V(\hat{\phi}), \\
  {\hT^\ptc{0}}_{~\ptc{i}} 
  &= -\frac{1}{N\sqrt{\gamma}}\hpi\partial_\ptc{i}\hat{\phi} .
\end{align}
By using the standard technique of the path integral, we have  
\begin{equation}
 \begin{split}
  \Tr e^{-\hK}&=\int d\phi\langle\phi|e^{- \hK}|\phi \rangle\\
  &  = 
  \int \mathcal{D}\phi \mathcal{D}\pi \exp\left(\int_0^{\beta_0} d\tau [i\int d^{d-1}\px  \partial_\tau\phi(\tau,\bm{\px})\pi(\tau,\bm{\px}) -\beta_0^{-1}K]\right),
  \label{eq:treK}
 \end{split}
\end{equation}
where $K$ denotes the functional corresponding to the operator $\hK$.
Parametrizing $\beta^\ptc{\mu} = \beta_0 e^{\sigma}u^\ptc{\mu}$
with the normalized vector $u^\mu$ satisfying $u^\mu u_\mu = - 1$
and $e^\sigma \equiv \sqrt{-\beta^\mu \beta_\mu }/\beta_0$,
and integrating Eq.~\eqref{eq:treK} 
with respect to the canonical momentum $\pi$,
we obtain the path-integral formula for the Massieu-Planck functional as
\begin{equation}
    \Psi[\pt; \lambda]
    =  \log\int \mathcal{D}\phi\, e^{S[\phi;\lambda]},
  \label{eq:RealScalarMPF}
\end{equation}
with
\begin{equation}
  \begin{split}
   S[\phi;\lambda] 
   &=\int_0^{\beta_{0}} d\tau\int d^{d-1}\px\sqrt{\gamma}\tilde{N} \Bigl[
   \frac{1}{2\tilde{N}^2}
   \Bigl(i\partial_\tau \phi -\tilde{N}^{\ptc{i}}\partial_{\ptc{i}}\phi \Bigr)^2
   -\Bigl(
   \frac{\gamma^{\ptc{i}\ptc{j}}}{2}\partial_{\ptc{i}}\phi\partial_{\ptc{j}}\phi+V(\phi)\Bigr) 
   \Bigr]\\
   &\equiv\int_0^{\beta_{0}} d^d \tilde{x} \sqrt{-\tilde{g}}\tilde{\mathcal{L}}(\phi, \tilde{\partial}_\ptc{\rho}\phi; \tilde{g}_{\ptc{\mu}\ptc{\nu}}),
  \end{split} 
  \label{eq:RealScalarAction}
\end{equation}
where we defined the Lapse function $\tilde{N}$, and
shift vector $\tilde{N}^\ptc{i}$ in the emergent thermal spacetime as
\begin{equation}
 \tilde{N}\equiv N e^{\dilatation} u^{\ptc{0}} =-n_\mu\beta^\mu/\beta_0, \quad 
  \tilde{N}^{\ptc{i}}\equiv \gamma^{\ptc{i}\ptc{j}}e^{\sigma}u_\ptc{j}
  = e^\sigma (u^\ptc{0}N^\ptc{i} + u^\ptc{i}) .
  \label{eq:tildeLapseShift}
\end{equation}
The partial derivative in the thermal space is defined as
$\tilde{\partial}_{\ptc{\mu}}\equiv (i\partial_{\tau},{\partial}_\ptc{i})$.
By using $\tilde{N}$ and $\tilde{N}^\ptc{i}$,
we also introduced the thermal metric $\tilde{g}_{\ptc{\mu}\ptc{\nu}}$
and its inverse $\tilde{g}^{\ptc{\mu}\ptc{\nu}}$ as
\begin{equation}
    \tilde{g}_{\ptc{\mu}\ptc{\nu}} 
    = \begin{pmatrix}
      -\tilde{N}^2 +\tilde{N}_\ptc{i}\tilde{N}^\ptc{i} & \tilde{N}_\ptc{j}\\
      \tilde{N}_{\ptc{i}}&  \gamma_{\ptc{i}\ptc{j}}
    \end{pmatrix},\qquad
    \tilde{g}^{\ptc{\mu}\ptc{\nu}}=
    \begin{pmatrix}
      -\tilde{N}^{-2}& \tilde{N}^{-2}\tilde{N}^{\ptc{j}}\\
      \tilde{N}^{-2} \tilde{N}^{\ptc{i}}&  \gamma^{\ptc{i}\ptc{j}}-\tilde{N}^{-2}\tilde{N}^{\ptc{i}}\tilde{N}^{\ptc{j}}
    \end{pmatrix}.
  \label{eq:thermalMetric}
\end{equation}
Here $\tilde{N}_\ptc{i} \equiv \gamma_{\ptc{i}\ptc{j}}\tilde{N}^\ptc{j}
= e^\sigma u_\ptc{i} $.
As is clearly demonstrated from Eqs.~\eqref{eq:RealScalarMPF} to \eqref{eq:thermalMetric},
the Masseiu-Planck functional $\Psi[\pt;\lambda]$ is expressed in terms of the path integral 
over the Euclidean action in the emergent curved spacetime, whose metric
is given in Eq.~\eqref{eq:thermalMetric} and determined
by the local thermodynamic parameters $\lambda^a(x)$. 
Note that in the hydrostatic gauge, we have 
$e^\sigma u^\ptc{\mu}= t^\ptc{\mu} = \delta^\ptc{\mu}_\ptc{0}$,
which leads to $\tilde{N} = N$ and $\tilde{N}^\ptc{i}=N^\ptc{i}$,
so that the thermal metric $\tilde{g}_{\ptc{\mu}\ptc{\nu}}$
coincides with the original metric $g_{\ptc{\mu}\ptc{\nu}}$
in our curved spacetime:
$\tilde{g}_{\ptc{\mu}\ptc{\nu}} = g_{\ptc{\mu}\ptc{\nu}}\big|_\mathrm{hs}$. 
We also note that the action takes imaginary values if 
$\tilde{N}^\ptc{i} \neq 0$, so that the lattice simulation may
suffer from the notorious sign problem in the presence of the
inhomogeneous fluid velocity.

\subsubsection{Charged scalar field}
We can easily generalize our analysis to a charged scalar field in a straightforward way
since we can decompose the charged scalar field into two real scalar fields.
This system, however, is distinct from simple summation
of two independent real fields in a sense that 
there exists a conserved charge curent coupled to the external gauge field, 
and thus has chemical potential. 
Dealing with the charged scalar field, 
we show how the chemical potential and external gauge field are implemented in our path-integral formula.

Lagrangian for a charged scalar boson is given by
\begin{equation}
\begin{split}
  \mathcal{L} 
  &= -g^{\ptc{\mu}\ptc{\nu}}D_\ptc{\mu}\Phi ^* D_\ptc{\nu}\Phi  - V(|\Phi| ^2) \\
  & =\frac{1}{N^2} (D_\ptc{t}\Phi ^* -N^\ptc{i} D_\ptc{i}\Phi^* )
  (D_\ptc{t}\Phi  -N^\ptc{i}D_\ptc{i}\Phi )
  - \gamma^{\ptc{i}\ptc{j}}D_{\ptc{i}}\Phi
  ^*D_{\ptc{j}}\Phi -V(|\Phi| ^2),
\end{split}
\end{equation}
where $\Phi$ denotes a complex field and describes bosons with positive and negative charges, 
and $D_\ptc{\mu}$ is a covariant derivative which acts charged fields as 
\begin{equation}
  D_\ptc{\mu} \Phi = \partial_\ptc{\mu}\Phi - i A_\ptc{\mu}\Phi, \quad
  D_\ptc{\mu} \Phi^* = \partial_\ptc{\mu}\Phi^* + i A_\ptc{\mu}\Phi^* .
\end{equation}
Here we take a charge of the complex field as unity, and $A_\ptc{\mu}$ denotes an external gauge field coupled to 
the conserved charge current.
Since this Lagrangian is invariant under $U(1)$ gauge transformation
$\Phi \rightarrow \Phi ' = e^{i\alpha}\Phi ,~
A_\mu \to A_\mu' = A_\mu + \partial_\mu \alpha$,
based on the general discussion given in Sec.~\ref{sec:Matter}, 
this system possesses the following conserved charge current, 
\begin{equation}
 J^\ptc{\mu} = -i g^{\ptc{\mu}\ptc{\nu}} (\Phi ^* D_\ptc{\nu}\Phi
  - \Phi D_\ptc{\nu} \Phi ^*).
\end{equation}

For the convenience, we decompose $\Phi$ into real and imaginary parts,
$\Phi = (\phi_1 + i\phi_2)/\sqrt{2}$, 
in which both of $\phi_1$ and $\phi_2$ denote real fields, 
and rewrite the Lagrangian in the following form,
\begin{equation}
  \begin{split}
    \mathcal{L} &= \sum_{a=1}^2\left[ -\frac{g^{\ptc{\mu}\ptc{\nu}}}{2} 
      D_\ptc{\mu}\phi_a D_\ptc{\nu}\phi_a  \right] - V(\phi_1^2
      + \phi_2 ^2) \\
    &= \sum_{a=1}^2 \left[
    \frac{1}{2N^2}(D_\pt\phi_a -N^{\ptc{i}}D_\ptc{i}\phi_a )^2
    - \frac{\gamma^{\ptc{i}\ptc{j}}}{2}D_{\ptc{i}}\phi_aD_{\ptc{j}}\phi_a \right]
    - V(\phi_1^2 + \phi_2 ^2).
  \end{split}
\end{equation}
Here we introduced a covariant derivative, which acts to real fields $\phi_a~(a=1,2)$ as
\begin{equation}
 D_\ptc{\mu} \phi_a
  = \partial_\ptc{\mu}\phi_a + \epsilon_{ab} A_\ptc{\mu} \phi_b ,
  \quad 
\end{equation}
with $\epsilon_{12} = 1 = -\epsilon_{21},~ \epsilon_{11}=\epsilon_{22}=0$, and 
use a contraction rule for the subscript $b$.

By using canonical momenta $\pi_a \equiv
- \sqrt{-g} g^{\ptc{0}\ptc{\nu}}D_\ptc{\nu} \phi_a 
=  N^{-1}\sqrt{\gamma}  ( D_\ptc{t} \phi_a - N^\ptc{i}D_\ptc{i}\phi_a)
\ (a=1,2)$, which satisfy the canonical commutation relations, 
$[\hat{\phi}_a (\pt,\bm{\px}), \hat{\pi}_b(\pt,\bm{\px}')]
= i\delta_{ab}\delta(\bm{\px}-\bm{\px}')$, 
all the conserved charge densities such as 
the energy-momentum and conserved $U(1)$ charge are written as 
\begin{align}
 {\hT^{\ptc{0}}}_{~\ptc{0}} 
 &= -\sum_{a=1}^2\left[ \frac{1}{2\gamma} \hpi_a^2
 + \frac{N^{\ptc{i}}}{N\sqrt{\gamma}} \hpi_a D_{\ptc{i}}\hat{\phi}_a
    + \frac{\gamma^{\ptc{i}\ptc{j}}}{2}D_{\ptc{i}}\hat{\phi}_a D_{\ptc{j}}\hat{\phi}_a \right]
  - V(\hat{\phi}_1^2 + \hat{\phi}_2^2) ,\\
  {\hT^\ptc{0}}_{~\ptc{i}} 
  &= -\sum_{a=1}^2 \frac{1}{N\sqrt{\gamma}} \hpi_a D_\ptc{i} \hat{\phi}_a ,\\
  \hat{J}^\ptc{0} 
  &=  \sum_{a=1}^2 \frac{1}{N\sqrt{\gamma}} \hpi_a \epsilon_{ab}\hat{\phi}_b .
\end{align}
Since these do not contain the time derivative
and the time component of the external gauge field $A_\ptc{0}$, 
these are invariant under the $U(1)$ gauge transformation
including $A_\ptc{0}$.

From this set of conserved quantities we obtain
\begin{equation}
  \Tr e^{-\hK}= \int \mathcal{D}\phi_1 \mathcal{D}\phi_2
  \mathcal{D}\pi_1 \mathcal{D}\pi_2\exp\left(\int_0^{\beta_0} d\tau 
    [i\int d^{d-1}\ptc{x} \sum_{a=1}^2 \partial_\tau \phi_a(\tau,\bm{\ptc{x}})\pi_a(\tau,\bm{\ptc{x}}) -\beta_0^{-1}K]\right).
\end{equation}
Since the canonical momenta are quadratic,
we are able to integrate out $\pi_a \ (a=1,2)$ also for this case.
Under the same parametrization $\beta^\ptc{\mu} = \beta_0 e^\sigma u^\ptc{\mu}$, we can write down 
the Masseiu-Planck functional as
\begin{equation}
 \Psi[\pt;\lambda, A_\ptc{\mu}] 
  = \log \int  \mathcal{D} \Phi \,
  e^{S[\Phi; \lambda,A_\ptc{\mu}]},
\end{equation}
with 
\begin{equation}
  \begin{split}
    S[\Phi;\lambda,A_\mu]
   &=\int_0^{\beta_{0}} d\tau\int d^{d-1} \px \sqrt{\gamma}\tilde{N} \\
   &\quad  \times\Biggl[\frac{1}{2\tilde{N}^2}
      \sum_{a=1}^2 \left[
   \Bigl(i\partial_\tau \phi_a + \epsilon_{ab} e^\sigma\mu \phi_b
   -\tilde{N}^{\ptc{i}}D_{\ptc{i}}\phi_a \Bigr)^2 
   - \frac{\gamma^{\ptc{i}\ptc{j}}}{2}
   {D}_{\ptc{i}}\phi_a {D}_{\ptc{j}}\phi_a\right]
   - V(|\Phi|^2)\Bigr) 
   \Biggr]\\ 
   &=  \int_0^{\beta_{0}} d^d \tilde{x} \sqrt{-\tilde{g}}
   \Biggl[
   \frac{1}{2\tilde{N}^2}
   \sum_{a=1}^2 \left[
   \Bigl(\tilde{D}_\ptc{0}\phi_a
   - \tilde{N}^{\ptc{i}} \tilde{D}_{\ptc{i}}\phi_a \Bigr)^2 
   - \frac{\gamma^{\ptc{i}\ptc{j}}}{2}
   \tilde{D}_{\ptc{i}}\phi_a \tilde{D}_{\ptc{j}}\phi_a\right]
      - V(|\Phi|^2)\Bigr) 
      \Biggr] \\ 
   &=\int_0^{\beta_{0}} d^d \tilde{x} \sqrt{-\tilde{g}}
   \Biggl[
   \frac{1}{\tilde{N}^2}
   \Bigl(\tilde{D}_\ptc{0}\Phi^*
   -\tilde{N}^{\ptc{i}}\tilde{D}_{\ptc{i}}\Phi^* \Bigr)
   \Bigl( \tilde{D}_\ptc{0} \Phi
   - \tilde{N}^{\ptc{i}} \tilde{D}_{\ptc{i}}\Phi\Bigr)
   -\gamma^{\ptc{i}\ptc{j}}{D}_{\ptc{i}}\Phi^* {D}_{\ptc{j}}\Phi 
   -V(|\Phi|^2)
   \Biggr]\\
   &\equiv\int_0^{\beta_{0}} d^d \tilde{x} \sqrt{-\tilde{g}}
    \tilde{\mathcal{L}}(\Phi, \tilde{\partial}_\ptc{\rho}\Phi; \tilde{g}_{\ptc{\mu}\ptc{\nu}},\tilde{A}_\ptc{\mu}),
  \end{split} 
\end{equation}
where we define the covariant derivative in thermal spacetime as follows:
\begin{align}
  \tilde{D}_\ptc{\mu} \phi_a &= \tilde{\partial}_\ptc{\mu} \phi_a + \epsilon_{ab}\tilde{A}_\ptc{\mu} \phi_b, 
  \label{eq:CovDer1}\\
  \tilde{D}_{\ptc{\mu}}\Phi &=\tilde{\partial}_\mu \Phi - {i}\tilde{A}_\ptc{\mu} \Phi, \quad
  \tilde{D}_\ptc{\mu} \Phi^* =  \tilde{\partial}_\mu\Phi^* + {i}\tilde{A}_\ptc{\mu}\Phi^* ,
  \label{eq:CovDer2}
\end{align}
with the external gauge field in thermal spacetime $\tilde{A}_\ptc{\mu}$
defined by
\begin{equation}
 \tilde{A}_\ptc{0} \equiv e^\sigma \mu = \nu /\beta_0, \quad
  \tilde{A}_\ptc{i}\equiv {A}_\ptc{i}.
  \label{eq:tildeA}
\end{equation} 
Here we note that $(i\partial_\tau)^\dag = i\partial_\tau$ in our convention.

We see that the resulting Euclidean action is again written in terms of 
the thermal metric background \eqref{eq:thermalMetric}, 
and an essential difference is only 
seen in the covariant derivative \eqref{eq:CovDer1} or \eqref{eq:CovDer2}. 
We, therefore, only need to consider the modified gauge connection 
in the presence of finite chemical potential, by replacing the 
partial derivative $\tilde{\partial}_\ptc{\mu}$ with the covariant one,
$\tilde{D}_\ptc{\mu}$. 
This brings about the gauge invariance in thermal spacetime.
As is discussed in Sec.~\ref{sec:Symmetry}, 
the additional term $e^\sigma \mu=\nu/\beta_0$ is
Kaluza-Klein gauge invariant, and thus,
the structure and symmetric properties of the emergent curved spacetime 
also hold for systems with finite chemical potential.

\subsection{Gauge field}
\subsubsection{Abelian gauge field}
As a next example, let us consider the electromagnetic field, 
whose field strength tensor is given by
\begin{equation}
  F_{\ptc{\mu}\ptc{\nu}} = \partial_\ptc{\mu} A_\ptc{\nu} - \partial_\ptc{\nu} A_\ptc{\mu},
\end{equation}
where $A_\ptc{\mu}$ denotes the four-vector potential. 
The Lagrangian for the electromagnetic field is
\begin{equation}
\begin{split}
  \mathcal{L} &= -\frac{1}{4}g^{\ptc{\mu}\ptc{\nu}}g^{\ptc{\alpha}\ptc{\beta}}F_{\ptc{\mu}\ptc{\alpha}}F_{\ptc{\nu}\ptc{\beta}} \\
  &= \frac{1}{2 N^2 }\gamma^{\ptc{i}\ptc{j}}
  ( F_{\ptc{0}\ptc{i}} - N^\ptc{k}F_{\ptc{k}\ptc{i}} )
  ( F_{\ptc{0}\ptc{j}} - N^\ptc{l}F_{\ptc{l}\ptc{j}} )
  -\frac{1}{4} \gamma^{\ptc{i}\ptc{j}}\gamma^{\ptc{k}\ptc{l}}
  F_{\ptc{i}\ptc{k}}F_{\ptc{j}\ptc{l}},
\end{split}
\end{equation}
where we use the coordinate system $(\pt, \bar{\bm{x}})$ with the ADM metric \eqref{eq:ADM} in the second line.

Since the field strength tensor is invariant under the gauge transformation
\begin{equation}
 A_\ptc{\mu}(x) \rightarrow A_\ptc{\mu}(x) + \partial_\ptc{\mu}\alpha(x),
\end{equation}
where $\alpha(x)$ is an arbitrary function smoothly dependent on $x$, 
the Lagrangian and all physical observables are also gauge invariant. 
However, to quantize gauge field in our setup, which is essentially Hamiltonian formalism, 
we need to fix a gauge. 
Here, we employ the axial gauge
\begin{equation}
 A_{\overline{d-1}} (x) = 0.
\end{equation}
Since this axial gauge condition does not completely fix the gauge, 
we fix the residual gauge freedom later on.

The canonical momenta $\Pi^\ptc{i}$ are now given by
\begin{equation}
 \Pi^\ptc{i} \equiv
  - \sqrt{-g}F^{\ptc{0}\ptc{i}}  
    =  \frac{\sqrt{\gamma}}{N}\gamma^{\ptc{i}\ptc{j}}(F_{\ptc{0}\ptc{j}} - N^\ptc{k}F_{\ptc{k}\ptc{j}})  .
\end{equation}
Note that $\Pi^\ptc{0}=0$, so that $A_\ptc{0}$ is not a dynamical field
because the field strength tensor is antisymmetric under the exchange of 
indices. 
We also note that due to the axial gauge condition $A_{\overline{d-1}}=0$, 
we do not have $\Pi^{\overline{d-1}}$ as a dynamical field. 
Indeed, it is determined by the Gauss's law
\begin{equation}
 \nabla_\ptc{i} F^{\ptc{0}\ptc{i}} = 0,
\label{eq:Gauss}
\end{equation}
where we consider the situation in the absence of the charged particles.

From the Lagrangian for the electromagnetic field,
we can construct energy-momentum tensor $\hT^{\mu}_{~\nu}$ as usual.  
It reads
\begin{align}
  {\hT^{\ptc{0}}}_{~\ptc{0}} &=
  \hat{F}^{\ptc{0}\ptc{\alpha}}\hat{F}_{\ptc{0}\ptc{\alpha}} + \hat{\mathcal{L}}
  =- \frac{1}{2\gamma}\hat{\Pi}^\ptc{i} \gamma_{\ptc{i}\ptc{j}} \hat{\Pi}^\ptc{j}
  -  \frac{N^\ptc{i}}{N\sqrt{\gamma}}\hat{F}_{\ptc{i}\ptc{j}} \hat{\Pi}^\ptc{j}
  -\frac{1}{4}\gamma^{\ptc{i}\ptc{j}}\gamma^{\ptc{k}\ptc{l}} \hat{F}_{\ptc{i}\ptc{k}} \hat{F}_{\ptc{j}\ptc{l}},
  \\
  {\hT^\ptc{0}}_{~\ptc{i}} &=
  \hat{F}^{\ptc{0}\ptc{\alpha}} \hat{F}_{\ptc{i}\ptc{\alpha}} 
  = - \frac{1}{N\sqrt{\gamma}}\hat{\Pi}^{\ptc{j}} \hat{F}_{\ptc{i}\ptc{j}}.
\end{align}
As is mentioned before, contrary to its apparent expression, $\Pi^{\overline{d-1}}$ is not 
an independent dynamical field, and determined by solving Gauss's law (\ref{eq:Gauss})
: $\Pi^{\overline{d-1}}=- \sqrt{-g} F^{\ptc{0}\, {\overline{d-1}}}(\Pi^\ptc{1},\cdots,\Pi^{\overline{d-2}})$. 
This fact is not useful in order to integrate out all the conjugate momentum $\Pi^\ptc{i}$. 
Therefore, we insert an identity 
\begin{equation}
  1 = \int \mathcal{D}\Pi^{\overline{d-1}} 
   \delta \big( \Pi^{\overline{d-1}} + \sqrt{-g}
   F^{\ptc{0}\,{\overline{d-1}}}(\Pi^\ptc{1},\cdots,\Pi^{\overline{d-2}} )\big)
\end{equation}
to avoid this apparent difficulty. 
Furthermore, by decomposing the Gauss law constraint as 
\begin{equation}
  \delta \big( \Pi^{\overline{d-1}} + \sqrt{-g}F^{\ptc{0}\, {\overline{d-1}}}(\Pi^\ptc{1},\cdots,\Pi^{\overline{d-2}}) \big)  
  =\delta ( \nabla_\ptc{i} \Pi^\ptc{i}) 
  \det \left( \frac{\partial  ( \nabla_\ptc{i} \Pi^\ptc{i}) }{\partial \Pi^{\overline{d-1}}}  \right)
  =\delta ( \nabla_\ptc{i} \Pi^\ptc{i}) \det \left( \nabla_{\overline{d-1}}  \right),
\end{equation}
we perform a similar analysis in the case of scalar fields, 
which results in the following path-integral expression:
\begin{equation}
  \begin{split}
    \Tr e^{-\hK}
    &= \int 
    \prod_{j=1}^{d-1} \mathcal{D}{\Pi}^\ptc{j}
    \prod_{k=0}^{d-2} \mathcal{D}{A}_\ptc{k}
    \, \delta (\nabla_\ptc{i} \Pi^\ptc{i} )  \det (\nabla_{\overline{d-1}}) \\
    & \quad\times \exp\left(\int_0^{\beta_0} d\tau \left[\int
	d^{d-1}\ptc{x}  \, 
	 \left( \sum_{l=1}^{d-2} \Pi^\ptc{l}i\partial_\tau A_\ptc{l} \right) 
	- \beta_0^{-1} K \right]\right) \\
    &= \int 
    \prod_{j=1}^{d-1} \mathcal{D}{\Pi}^\ptc{j}
    \prod_{k=0}^{d-2} \mathcal{D}{A}_\ptc{k}
    \, \det (\nabla_{\overline{d-1}}) \\
    & \quad\times \exp\left(\int_0^{\beta_0} d\tau \left[ \int
	d^{d-1}\ptc{x}  \, 
	 \left( \sum_{l=1}^{d-2} \Pi^\ptc{l}i\partial_\tau A_\ptc{l} 
	 - \Pi^\ptc{i} i \partial_\ptc{i} A_\ptc{0} \right)
      - \beta_0^{-1} K\right] \right),
   \label{eq:PathIntegral-gauge}
  \end{split}
\end{equation}
where we used a functional-integral expression for the delta function $\delta(\nabla_\ptc{i}\Pi^\ptc{i})$ 
with an auxiliary field $A_\ptc{0}$
\begin{equation}
  \delta(\nabla_\ptc{i}\Pi^\ptc{i}) = \int \mathcal{D} A_\ptc{0} 
  \exp \left( i\int_0^{\beta_0} d\tau \int d^{d-1} \bar{x} 
    A_\ptc{0} \nabla_\ptc{i}\Pi^\ptc{i}  \right),
\end{equation}
and perform an integration by parts in order to obtain the last line in Eq.~(\ref{eq:PathIntegral-gauge}).

Using the same parametrization
$\beta^\ptc{\mu} = \beta_0 e^\sigma u^\ptc{\mu}$
as the case of the scalar fields, 
and after integrating out the conjugate momenta $\Pi^\ptc{i}$, we obtain the path-integral formula for 
the Masseiu-Planck functional,
\begin{equation}
  \Psi[\pt;\lambda] = \log \int 
  \prod_{i=0}^{d-2}\mathcal{D} A_\ptc{i}
  \det (\nabla_{\overline{d-1}})
  \exp \left( {S[A_\ptc{\sigma}; \lambda]} \right),
\end{equation}
with
\begin{equation}
  \begin{split}
    S[A_\ptc{\sigma};\lambda] 
   &=\int_0^{\beta_{0}} d\tau\int d^{d-1}\px\sqrt{\gamma}\tilde{N}
   \Bigl[        \frac{1}{2\tilde{N}^2}\gamma^{\ptc{i}\ptc{j}}
      ( \tilde{F}_{\ptc{0}\ptc{i}} - \tilde{N}^\ptc{k}F_{\ptc{k}\ptc{i}} )
      ( \tilde{F}_{\ptc{0}\ptc{j}} - \tilde{N}^\ptc{l}F_{\ptc{l}\ptc{j}} )
      -\frac{1}{4} \gamma^{\ptc{i}\ptc{j}}\gamma^{\ptc{k}\ptc{l}}
      F_{\ptc{i}\ptc{k}}F_{\ptc{j}\ptc{l}} \Bigr],\\
    &\equiv\int_0^{\beta_{0}} d^d \tilde{x} \sqrt{-\tilde{g}}
    \tilde{\mathcal{L}}(\tilde{\partial}_\ptc{\rho}A_\ptc{\sigma}; \tilde{g}_{\ptc{\mu}\ptc{\nu}}),
  \end{split} 
\end{equation}
where $\tilde{N}$ and $\tilde{N}^{\ptc{i}}$ are defined
in Eq.~\eqref{eq:tildeLapseShift}. 
It should be emphasized that they are completely same as the case for
the scalar fields, and thus, the thermal metric
$\tilde{g}_{\ptc{\mu}\ptc{\nu}}$ is expected to be universal regardless of
the spin of microscopic quantum fields. 
Here we introduced the field strength tensor along the imaginary-time direction:
\begin{equation}
 \tilde{F}_{\ptc{0}\ptc{i}}
  \equiv i\partial_\tau A_\ptc{i} - i\partial_\ptc{i} A_\ptc{0} .
\end{equation}
The most important point is that the result is again written
in terms of the Euclidean action in the emergent curved spacetime 
with the thermal metric \eqref{eq:thermalMetric}.

A short comment on the gauge invariance is in order here. 
The above result is the path-integral formula of the Massieu-Planck functional for the axial gauge, and 
the path integral over $A_{\overline{d-1}}$ is not contained because of the axial gauge condition $A_{\overline{d-1}} =0$. 
However, we can implement the axial gauge condition through an insertion of 
\begin{equation}
 1 = \int \mathcal{D} A_{\overline{d-1}} \delta(A_{\overline{d-1}}),
\end{equation}
and, as a result, we obtain 
\begin{equation}
  \Psi[\pt;\lambda] = \log \int \mathcal{D}A_\ptc{\mu}
   \delta(A_{\overline{d-1}}) \det (\nabla_{\overline{d-1}})
  e^{S[ A_\ptc{\sigma}; \lambda]}.
\end{equation}
This is the result for a special choice of the axial gauge, 
but we can easily generalize this result for an arbitrary gauge choice 
by replacing the gauge fixing condition and Jacobian as 
\begin{equation}
   \delta(A_{\overline{d-1}}) \det (\nabla_{\overline{d-1}}) \rightarrow  \delta(F) \det \left( \frac{\partial F}{\partial \alpha} \right),
\end{equation}
where $F=0$ gives the gauge fixing condition like $F = A_{\overline{d-1}}$ in the axial gauge. 
Since the delta function and the determinant give a gauge-invariant combination, 
the final expression for the Masseiu-Planck functional is given by
\begin{equation}
  \Psi[\pt;\lambda] = \log \int \mathcal{D}A_\ptc{\mu}
   \delta(F) \det \left( \frac{\partial F}{\partial \alpha} \right)
  \exp \left({S[ A_\ptc{\sigma}; \lambda]} \right) , 
\end{equation}
which is explicitly gauge invariant, so that 
we can choose an arbitrary gauge suitable for our calculation.

\subsubsection{Non-Abelian gauge field}
Let us generalize our result to the non-Abelian gauge field. 
Here, for concreteness, we consider $SU(N)$ gauge theory.
The Lagrangian for the non-abelian gauge field is given by
\begin{equation}
  \begin{split}
  \mathcal{L} &= -\frac{1}{4}g^{\ptc{\mu}\ptc{\nu}}g^{\ptc{\alpha}\ptc{\beta}}{G^a}_{\ptc{\mu}\ptc{\alpha}}{G^a}_{\ptc{\nu}\ptc{\beta}} \\
  &= \frac{1}{2N^2}\gamma^{\ptc{i}\ptc{j}}
  ( G^a_{~\ptc{0}\ptc{i}} - N^\ptc{k}G^a_{~\ptc{k}\ptc{i}} )
  ( G^a_{~\ptc{0}\ptc{j}} - N^\ptc{l}G^a_{~\ptc{l}\ptc{j}} )
  -\frac{1}{4} \gamma^{\ptc{i}\ptc{j}}\gamma^{\ptc{k}\ptc{l}}
  G^a_{~\ptc{i}\ptc{k}}G^a_{~\ptc{j}\ptc{l}}.
\end{split}
\label{eq:nonAbelian}
\end{equation}
Here we introduced the field strength tensor for the non-Abelian gauge field,
\begin{equation}
  {G^a}_{\ptc{\mu}\ptc{\nu}} = \partial_\ptc{\mu}{A^a}_\ptc{\nu}-\partial_\ptc{\nu}{A^a}_\ptc{\mu} 
  + gf_{abc}{A^b}_\ptc{\mu}{A^c}_\ptc{\nu},
\end{equation}
with the non-Abelian gauge field $A^a_{~\ptc{\mu}}$, the dimensionless coupling constant $g$, 
and the structure constants of $SU(N)$ gauge group $f_{abc}$,
which satisfy
\begin{equation}
 [t^a, t^b] = i f_{abc} , \quad \mathrm{tr}(t^at ^b)= \frac{1}{2}\delta_{ab}, 
\end{equation}  
where $t^a$ denotes generators of $SU(N)$ group.
Note that the summation over repeated indices is assumed. 
One important difference with the Abelian gauge field is that the gauge field carries the (color) index $a$ which runs 
from $a=1$ to $N^2-1$. 
Introducing $A_\ptc{\mu} =t^aA^a_{~\ptc{\mu}}$, we can express the field strength tensor 
in terms of the commutator of the covariant derivative:
\begin{equation}
  G_{\ptc{\mu}\ptc{\nu}} 
  =\partial_\ptc{\mu} A_\ptc{\nu}
  - \partial_\ptc{\nu} A_\ptc{\mu} - ig[A_\ptc{\mu}, A_\ptc{\nu}] 
  = \frac{i}{g}[D_\ptc{\mu}, D_\ptc{\nu}],
\end{equation} 
where we introduced $G_{\ptc{\mu}\ptc{\nu}} \equiv t^a
G^a_{~{\ptc{\mu}\ptc{\nu}}}$ and covariant derivative, 
\begin{equation}
 D_\ptc{\mu} \equiv \partial_\ptc{\mu} - i g t^a A^a_{~\ptc{\mu}}. 
\end{equation}
The field strength tensor transforms as $G_{\ptc{\mu}\ptc{\nu}} \rightarrow U G_{\ptc{\mu}\ptc{\nu}} U^\dag$
under the $SU(N)$ gauge transformation
\begin{equation}
  A_\ptc{\mu}(x) \rightarrow U(x) (A_\ptc{\mu}(x) + ig^{-1}\partial_\ptc{\mu})U^\dag (x),
\end{equation}
where $U(x) \equiv \exp(  i\theta^a(x)t^a)$ is a unitary matrix:
$UU^\dag = \mathbb{1}$.
Together with the cyclic property of traces: $\Tr \left(AB\right) =\Tr \left(BA\right)$, 
we can easily see gauge invariance of the Lagrangian \eqref{eq:nonAbelian}.

Quantization procedure of the non-Abelian gauge field is 
accomplished in a similar way with the Abelian gauge fields, 
and we directly write down the final result for the Masseiu-Planck functional,
\begin{equation}
  \Psi[\pt;\lambda] = \log \int \mathcal{D} A^a_{~\ptc{\mu}} 
  \delta (F^b) \det \left( \frac{\partial F^c}{\partial \alpha_d}  \right) 
  e^{S[A_\ptc{\sigma};\lambda]},
\end{equation}
where $\delta(F^b)$ represents the gauge-fixing condition, and the determinant does 
the Fadeev-Popov determinant with the gauge parameter $\alpha^d$.
The resulting Euclidean action is the completely same as the previous
analysis on the Abelian case, which reads
\begin{equation}
  \begin{split}
   S[A_\ptc{\sigma};\lambda] 
   &=\int_0^{\beta_{0}} d\tau\int d^{d-1}\px\sqrt{\gamma}\tilde{N} \Bigl[
   \frac{1}{2\tilde{N}^2}\gamma^{\ptc{i}\ptc{j}}
   ( G^a_{~\tau\ptc{i}} - \tilde{N}^\ptc{k}G^a_{~\ptc{k}\ptc{i}} )
   ( G^a_{~\tau\ptc{j}} - \tilde{N}^\ptc{l}G^a_{~\ptc{l}\ptc{j}} )
   -\frac{1}{4} \gamma^{\ptc{i}\ptc{j}}\gamma^{\ptc{k}\ptc{l}}
   G^a_{~\ptc{i}\ptc{k}}G^a_{~\ptc{j}\ptc{l}} \Bigr],\\
   &\equiv
   \int_0^{\beta_{0}} d^d \tilde{x} \sqrt{-\tilde{g}}
   \tilde{\mathcal{L}}(\tilde{\partial}_\ptc{\rho}A_\ptc{\sigma}; \tilde{g}_{\ptc{\mu}\ptc{\nu}}).
  \end{split} 
\end{equation}
Here in the same way as the Abelian case, we introduced the field strength tensor along the imaginary-time direction:
\begin{equation}
 \tilde{G}^a_{~\ptc{0}\ptc{i}} \equiv i\partial_\tau A^a_{~\ptc{i}} - i\partial_\ptc{i} A^a_{~\ptc{0}}.
\end{equation}
The result is again written in terms of the path integral of the
Euclidean action in the thermally emergent curved spacetime
with the same thermal metric defined in Eq.~\eqref{eq:thermalMetric}.

\subsection{Dirac field}
\label{sec:Dirac}
\subsubsection{Spinor field in curved spacetime}
As the last example, let us consider the Dirac field. 
Before starting the path-integral analysis on the Dirac field,
we first summarize a way to describe spinor fields in the curved spacetime, 
that is, the so-called vielbein formalism
(See e.g. \cite{ParkerToms} in more detail).

In order to describe the spinor field in the curved spacetime,
we use the vielbein $e_\mu^{~a}$ 
instead of the metric $g_{\mu\nu}$. 
Here, Greek letters ($\mu, \nu,\cdots$) represent the curved spacetime indices in the coordinate system $(t,\bm{x})$,
while Latin letters ($a, b, \cdots$) do the local Lorentz indices.
The metric and vielbein are related to each other through
\begin{equation}
  g_{\mu\nu}=e_\mu^{~a} e_\nu^{~b} \eta_{ab}, \quad
  \eta^{ab}=e_\mu^{~a} e_\nu^{~b} g^{\mu\nu}.
  \label{eq:MetricVielbein}
\end{equation}
We also define the inverse vielbein $e_a^{~\mu}$, which satisfies the relations 
$\delta_{\mu}^\nu=e_\mu^{~a}e^{~\nu}_{a}$, $\delta_{a}^b= e^{~\mu}_a e_{\mu}^{~b}$.
The (inverse) vielbein enables us to exchange the curved spacetime indices and the local Lorentz indices as follows:
\begin{equation}
  \begin{split}
    B_a &= e_{a}^{~\mu} B_\mu, \quad B^a = e^{~a}_\mu B^\mu, \\
    B_\mu &= e_{\mu}^{~a}B_a, \quad B^\mu = e_a^{~\mu} B^a.
  \end{split}
\end{equation}

The Lagrangian for the Dirac field $\psi$ is expressed by the use of the inverse vielbein
\begin{equation}
  \begin{split}
  \mathcal{L} 
   &=-\frac{1}{2}\ptc{\psi}
   (e_a^{~\mu}\gamma^a\ra{D}_\mu -\la{D}_\mu  e_a^{~\mu}\gamma^a)\psi  
  -m\ptc{\psi}\psi,
  \label{eq:DiracLagrangian}
  \end{split}
\end{equation}
where we defined $\ptc{\psi}\equiv i\psi^\dag\gamma^0$, and the covariant derivative 
\begin{equation}
  D_\mu=\partial_\mu-i (\cA_\mu + A_\mu) 
  \quad
  \mathrm{with}
  \quad
  \cA_\mu=\frac{1}{2} \omega_\mu^{~ab} \varSigma_{ab},
  \label{eq:CovDerFermion}
\end{equation}
with the external gauge field $A_\mu$.
Here $\varSigma_{ab} \equiv i[\gamma_a,\gamma_b]/4$ is a
generator of the Lorentz group with $\gamma^a$ being the 
gamma matrices, which satisfy a set of relations 
$\{\gamma^a,\gamma^b\}= 2\eta^{ab}$ with $\{A,B \} \equiv AB + BA$, 
$(i\gamma^0)^\dag=i\gamma^0$, 
$(i\gamma^0)^\dag (i\gamma^0)=1$,
$i\gamma^0 (\gamma^a)^\dag i\gamma^0=-\gamma^a$, 
and $i\gamma^0\varSigma_{ab}^\dag i\gamma^0=\varSigma_{ab}$ in our convention. 
From a direct calculation, we can check the following relations:
\begin{align}
  [\varSigma_{ab},\varSigma_{cd} ]
  &= -i(\eta_{ac}\varSigma_{bd}-\eta_{nc}\varSigma_{ad}
    -\eta_{ad}\varSigma_{bc}+\eta_{bd}\varSigma_{ac}),\\
  [\gamma_a,\varSigma_{bc} ]&= -i(\eta_{ac}\gamma_b - \eta_{ab}\gamma_c).
\end{align}
The left and right derivatives are defined as
\begin{equation}
 \ra{D}_\mu \phi \equiv \partial_\mu\phi -i\mathcal{A}_\mu\phi
  \quad\text{and}\quad
    \phi\la{D}_\mu  \equiv \partial_\mu\phi +i\phi\mathcal{A}_\mu.
\end{equation}
In Eq.~\eqref{eq:CovDerFermion}, we have a spin connection $\omega_\mu^{~ab}$,
which is expressed by the vielbein as 
\begin{align}
  \omega_\mu^{~ab} 
  &=\frac{1}{2}e^{a\nu}e^{b\rho} 
  (C_{\nu\rho\mu}-C_{\rho\nu\mu}-C_{\mu\nu\rho}), \\
  C_{\mu\nu\rho}
  &\equiv e_{\mu}^{~c}(\partial_{\nu}e_{\rho c}-\partial_{\rho}e_{\nu c}),
\end{align}
where $C_{\mu\nu\rho}$ are called the Ricci rotation coefficients. 
Here we assume the torsion-free condition for the background curved spacetime. 
We note that the spin connection $\omega_\mu^{~ab}$ is
anti-symmetric under the exchange of 
the local Lorentz indices: $\omega_\mu^{~ab} = - \omega_\mu^{~ba}$.

\subsubsection{Energy-momentum conservation law for spinor field}
As is demonstrated in Sec.~\ref{sec:Matter}, taking the variation of the action with respect to the metric, 
we obtain the conserved energy-momentum tensor associated with 
diffeomorphism invariance.
However, if matters considered are composed of spinor fields, 
the action is described not by the metric $g_{\mu\nu}$
but by the vielbein $e_\mu^{~a}$ as 
\begin{equation}
  S[\psi,\ptc{\psi}; e_\mu^{~a}, A_\mu] = \int d^d x e 
  \mathcal{L}(\psi(x),\ptc{\psi}(x), D_\mu \psi (x), D_\mu \ptc{\psi} (x);
  e_\mu^{~a}(x), A_\mu(x)) ,
\end{equation} 
where we define $e \equiv \det e_\mu^{~a} = \sqrt{-g}$, and 
the explicit form of the Lagrangian for the Dirac field is already given by Eq.~\eqref{eq:DiracLagrangian}.
In a similar way discussed in Sec.~\ref{sec:Matter},
we can generalize our discussion on the derivation of the energy-momentum conservation law for the fermionic action.
Let us consider a set of variations with respect to the vielbein $e_\mu^{~a}$, 
the external gauge field $A_\mu$, and the spinor fields $\psi,~\ptc{\psi}$:
\begin{align}
  \lie_\xi e_\mu^{~a} &\equiv {e'}_\mu^{~a}(x) - e_\mu^{~a} (x)
  =  \xi^\nu \nabla_\nu e_\mu^{~a} + e_\nu^{~a} \nabla_\mu \xi^\nu, \\
  \lie_\xi A_\mu &\equiv A_\mu' (x) - A_\mu (x) = \xi^\nu \nabla_\nu A_\mu 
  + A_\nu \nabla_\mu \xi^\nu, \\
  \lie_\xi \psi &\equiv \psi' (x) - \psi (x) =  \xi^\nu \partial_\nu \psi, \\
  \lie_\xi \ptc{\psi} &\equiv \ptc{\psi}' (x) - \ptc{\psi} (x) =  \xi^\nu \partial_\nu \ptc{\psi},
\end{align}
which are caused by the general coordinate transformation \eqref{eq:Diffeo}. 
Because the action again has diffeomorphism invariance,
the variation of the action under this transformation vanishes:
$\delta S =0$.
Furthermore, since the variations of the fields does not
contribute with the help of the equation of motion, 
the variation of the action leads to
\begin{equation}
  \begin{split}
   \delta S 
    &= \int d^d x e
    \left[  \mathcal{T}^\mu_{~a} \lie_\xi e_\mu^{~a}
      + J^\mu \lie_\xi A_\mu 
    \right] \\
    &= \int d^d x e
   \left[ \mathcal{T}^\mu_{~a}
   (\xi^\nu \nabla_\nu e_\mu^{~a} + e_\nu^{~a} \nabla_\mu \xi^\nu)
      + J^\mu (\xi^\nu \nabla_\nu A_\mu + A_\nu\nabla_\mu \xi^\nu) 
    \right] \\
    &= -\int d^d x e
   \left[
   (\nabla_\mu \mathcal{T}^{\mu}_{~\nu} - F_{\nu\lambda} J^\lambda ) \xi^\nu
   \right]  
   + \int d^d x e  \mathcal{T}_{ab} \omega_\nu^{ab} \xi^\nu \\
    &\quad + \int d^d x e\nabla_\mu[(\mathcal{T}^\mu_{~\nu} + J^\mu A_\nu)\xi^\nu]
    - \int d^d x e \xi^\nu A_\nu \nabla_\mu J^\mu ,
    \label{eq:deltaS2-2}
  \end{split}
\end{equation}
where we define the energy-momentum tensor for spinor fields
$\mathcal{T}^\mu_{~a}$ as
\begin{equation}
 \mathcal{T}^\mu_{~a} \equiv \frac{1}{e} \frac{\delta S}{\delta e_\mu^{~a}},
   \label{eq:EMFermion1}
\end{equation}
and we replace the Lorentz indices as the curved spacetime indices
by the use of the vielbein: 
$\mathcal{T}^\mu_{~\nu} = e_\nu^{~a}\mathcal{T}^\mu_{~a}$. 
Here we also used a so-called tetrad postulate that the covariant derivative of 
the vielbein vanishes:
\begin{equation}
  D_\mu e_\nu^{~a} = \nabla_\mu e_\nu^{~a} 
  + \omega_{\mu~b}^{~a} e_\nu^{~b} = 0,
\end{equation}
where $\nabla_\mu e_\nu^{~a} \equiv \partial_\mu e_\nu^{~a}
- \Gamma^{\rho}_{~\mu\nu} e_\rho^{~a} $
with the usual Christoffel symbol $\Gamma^\rho_{~\mu\nu}$. 
Compared to the previous case in Eq.~\eqref{eq:deltaS}, 
we have the additional term proportional to $\mathcal{T}_{ab} \omega_\mu^{~ab}$, 
which, in general, does not seem to vanish.
However, as will be shown soon, this term vanishes
due to local Lorentz invariance,
and we obtain the energy-momentum conservation law 
\begin{equation}
  \nabla_\mu \mathcal{T}^\mu_{~\nu} = F_{\nu\lambda} J^\lambda.
  \label{eq:EMConservationFermion}
\end{equation}

Let us focus on the reason that the additional term does not contribute.
In addition to diffeomorphism invariance, we have another symmetry 
due to the fact that it does not matter which locally inertial frames we adopt.
In other words, the fermionic action is invariant under the local Lorentz transformation:
\begin{align}
  \delta_\alpha e_\mu^{~a} &= \alpha^a_{~b}(x) e_\mu^{~b}, \\
  \delta_\alpha \psi &= -\frac{i}{2}\alpha^{ab}(x) \varSigma_{ab} \psi , \\
  \delta_\alpha \ptc{\psi} &= \frac{i}{2} \alpha^{ab}(x) \ptc{\psi}\varSigma_{ab}  , 
\end{align}
where $\alpha^a_{~b} (x)$ denotes a local rotation angle,
which is anti-symmetric: $\alpha_{ab} (x)= - \alpha_{ba}(x)$, 
and $\varSigma_{ab}$ the generator of the Lorentz group.
By the use of the equation of motion $\delta S/ \delta\psi = \delta S / \delta \ptc{\psi}=0$, 
the variation of the action under the infinitesimal
local Lorentz transformation is expressed as
\begin{equation}
  \begin{split}
   \delta S &= \int d^d x e \mathcal{T}^\mu_{~a} \delta_\alpha e_\mu^{~a} 
   = -\int d^d x e \mathcal{T}^{ab} \alpha_{ab} \\
   &= -\int d^d x e \frac{1}{2} (\mathcal{T}^{ab} - \mathcal{T}^{ba} )
   \alpha_{ab} ,
  \end{split} 
\end{equation}
for arbitrary $\alpha_{ab}(x)$. 
Therefore, local Lorentz invariance of the action: $\delta S=0$, results in the proposition 
that the anti-symmetric part of the energy-momentum tensor vanishes:
\begin{equation}
  \mathcal{T}^{ab} - \mathcal{T}^{ba} = 0 .
  \label{eq:EMSymmetry}
\end{equation}
This is the reason why we drop the term proportional to
$\mathcal{T}_{ab}\omega_\mu^{~ab}$ in Eq.~\eqref{eq:deltaS2-2}.

Combined with the consequence of differmorphism invariance and that of local Lorentz invariance, 
in other words, the energy-momentum conservation law \eqref{eq:EMConservationFermion}, 
and the symmetric property of the energy-momentum tensor \eqref{eq:EMSymmetry},
we immediately conclude that the symmetric energy-momentum tensor 
is also conserved for the fermionic case,
\begin{equation}
  \nabla_\mu T^{\mu}_{~\nu} = F_{\nu\lambda}J^\lambda,
\end{equation}
where we define the symmetric energy-momentum tensor $T^{\mu\nu}$ as
\begin{equation}
 T^{\mu\nu} = \frac{1}{2}
  ( \mathcal{T}^\mu_{~a} e^{\nu a} + \mathcal{T}^\nu_{~a} e^{\mu a}  ),
  \label{eq:EMFermion2}
\end{equation}
which is clearly symmetric under $\mu \leftrightarrow \nu$ by definition.

\subsubsection{Charge conservation law for spinor field} 
The Lagrangian for the Dirac field \eqref{eq:DiracLagrangian}
also has a gauge symmetry under the $U(1)$ gauge transformation defined in
Eqs.~\eqref{eq:GaugeTr1} and \eqref{eq:GaugeTr2}. 
We, therefore, have a conserved vector current $J^\mu$
which is coupled to the background $U(1)$ gauge field,
\begin{equation}
  J^\mu \equiv i\ptc{\psi}\gamma^\mu \psi
  \quad
  \mathrm{with}
  \quad 
  \nabla_\mu J^\mu = 0.
\end{equation}


\subsubsection{Path-integral formula for Dirac field}
We are ready to develop the path-integral formula for the Dirac field. 
First of all, taking the variation of the action with respect to vielbein, 
we obtain the energy-momentum tensor $\mathcal{T}^{\mu\nu}$ defined in Eq.~\eqref{eq:EMFermion1} as
\begin{equation}
 \mathcal{T}^{\mu\nu}
  = \frac{1}{2}\ptc{\psi} (\gamma^{\mu}\ra{D}^\nu-\la{D}^\nu \gamma^\mu )\psi
  -\frac{i}{4}D_\rho \big(\ptc{\psi} \{\gamma^\mu,\Sigma^{\nu\rho}\}\psi\big)
  + g^{\mu\nu}\mathcal{L} .
  \label{eq:EMFermion3}
\end{equation}

By symmetrizing the indices, we also have the symmetric energy-momentum tensor $T^{\mu\nu}$ defined in Eq.~\eqref{eq:EMFermion2}
\begin{equation}
    T^{\mu\nu} 
    \equiv \frac{1}{2}(\mathcal{T}^{\mu\nu} + \mathcal{T}^{\nu\mu})
    = \frac{1}{4}\ptc{\psi} (\gamma^{\mu}\ra{D}^\nu + \gamma^{\nu}\ra{D}^\mu
      -\la{D}^\nu \gamma^\mu -\la{D}^\mu \gamma^\nu)\psi
     +g^{\mu\nu}\mathcal{L} .
    \label{eq:EMFermion4}
\end{equation}

It is worth to clarify which energy-momentum tensor
we adopt in order to construct the local Gibbs distribution. 
Our choice is the symmetric energy-momentum tensor \eqref{eq:EMFermion4}%
\footnote{
Of course, we can choose Eq.~\eqref{eq:EMFermion3} together
with the condition \eqref{eq:EMSymmetry} originated from
local Lorentz invariance. 
These choices are equivalent. 
}
not the so-called canonical energy-momentum tensor. 
The reason for this choice is answered from several viewpoints as follows: 
First, we do not have the global translational symmetry
in the presence of the background fields, and do not have
the corresponding Noether current, or the canonical energy-momentum tensor. 
Second, our guiding principle to construct the local Gibbs distribution 
is that we should collect a set of independent conserved quantities such as the energy, momentum, and conserved charge. 
We do not have to take into account the angular momentum as a conserved charge 
since if the energy-momentum tensor is symmetric, the associated angular momentum is trivially conserved, 
and hence, it is not the independent conserved quantity. 
To avoid question that we should consider the angular momentum or not, 
we choose the symmetric energy-momentum tensor. 
One can also argue that the energy-momentum tensor appeared in
general relativity should be symmetric, 
and we should also choose the symmetric one
to discuss relativistic hydrodynamics.

If we adopt the symmetric energy-momentum tensor, we have 
\begin{equation}
 \Tr e^{-\hK} = \int \mathcal{D}\psi\mathcal{D}\bar{\psi}
  \exp\left(\int_0^{\beta_0} d\tau \left[i\int
       d^{d-1}\ptc{x} e \frac{-1}{2}
       ( \ptc{\psi}\gamma^\ptc{0} \ra{\partial_\tau}{\psi}
       - \ptc{\psi} \la{\partial_\tau}\gamma^\ptc{0}{\psi}) 
       - \beta_0^{-1}K \right]\right), 
\end{equation}
where $K$ includes the symmetric energy-momentum tensor. 
Here we note that the imaginary-time derivative is not the covariant derivative but the partial derivative, 
because it simply arises from inner products of the adjacent state vectors introduced by the insertion of complete sets. 
On the other hand, the spatial derivative is the covariant derivative
whose spin connection is composed of the vielbein $e_\mu^{~a}$.
Nevertheless, note that it is not trivial that this spin connection gives 
the correct one for the emergent thermal spacetime, in which
one direction is not real time but imaginary time. 
We will show that they coincide with each other later on. 

Contrary to the previous examples, we face with the problematic situation that 
the symmetric energy-momentum tensor does not seem to reproduce the correct Euclidean action. 
It is also not reasonable that the imaginary-time derivative is not covariant one, 
if the Euclidean action is given as that in the emergent curved spacetime. 
As will be shown below, these difficulties are closely related with each other, 
and a proper treatment again gives the correct Euclidean action in the emergent thermal spacetime.

In order to decompose the symmetric energy-momentum tensor, we use the consequence of local Lorentz invariance
\begin{equation}
    \mathcal{T}^{mn} - \mathcal{T}^{nm} =0  
     ~\Leftrightarrow~
    \frac{1}{4} \ptc{\psi} (\gamma^{\mu}\ra{D}_\nu - \gamma_{\nu}\ra{D}^\mu
      -\la{D}_\nu \gamma^\mu + \la{D}^\mu \gamma_\nu)\psi
    - \frac{i}{4} D_\rho( \ptc{\psi}\{\gamma^\mu, \Sigma_\nu^{~\rho} \} \psi) = 0.
\end{equation}
By the virtue of this relation, we can rewrite the symmetric energy-momentum tensor as 
\begin{equation}
  \begin{split}
    T^\mu_{~\nu}
    &= \frac{1}{2}\ptc{\psi} (\gamma^{\mu}\ra{D}_\nu-\la{D}_\nu \gamma^\mu )\psi
    + \delta^\mu_\nu \mathcal{L}
    - \frac{1}{4} \ptc{\psi} (\gamma^{\mu}\ra{D}_\nu - \gamma_{\nu}\ra{D}^\mu
      -\la{D}_\nu \gamma^\mu + \la{D}^\mu \gamma_\nu)\psi \\
    &= \frac{1}{2}\ptc{\psi} (\gamma^{\mu}\ra{D}_\nu-\la{D}_\nu \gamma^\mu )\psi
    + \delta^\mu_\nu \mathcal{L}
    - \frac{i}{4} D_\rho( \ptc{\psi}\{\gamma^\mu, \Sigma_\nu^{~\rho} \} \psi) \\
   &\equiv \Theta^\mu_{~\nu}
   - \frac{1}{2} D_\rho \Sigma^{\mu~\rho}_{~\nu},
  \end{split}
\end{equation}
where we defined the canonical part of the energy-momentum tensor $\Theta^\mu_{~\nu}$, 
and the spin part of the angular momentum tensor $\Sigma^{\mu~\rho}_{~\nu}$ as
\begin{align}
  \Theta^\mu_{~\nu} 
 &\equiv \frac{1}{2}\ptc{\psi}
 (\gamma^{\mu}\ra{D}_\nu-\la{D}_\nu \gamma^\mu )\psi
  + \delta^\mu_\nu \mathcal{L}, \\
 \Sigma^{\mu~\rho}_{~\nu} &\equiv
 \frac{i}{2}
  \ptc{\psi}\{\gamma^\mu, \Sigma_\nu^{~\rho} \} \psi.
 \label{eq:SpinAM}
\end{align}

Then, we can rewrite $K$ as follows:
\begin{equation}
  \begin{split}
   K
   &= -\int d \Sigma_{\pt \ptc{\mu}} 
   \left( \beta^\ptc{\nu}T^{\ptc{\mu}}_{~\ptc{\nu}}
   + \nu J^\ptc{\mu} \right) \\
   &= -\int d \Sigma_{\pt \ptc{\mu}} 
    \left( \beta^\ptc{\nu} \Theta^{\ptc{\mu}}_{~\ptc{\nu}}
      + \frac{1}{2} \Sigma^{\ptc{\mu}~\ptc{\rho}}_{~\ptc{\nu}} D_\ptc{\rho} \beta^\ptc{\nu} 
   + \nu J^\ptc{\mu} \right)
   + \frac{1}{2} \int d \Sigma_{\pt \ptc{\mu}} D_\ptc{\rho}( \Sigma^{\ptc{\mu}~\ptc{\rho}}_{~\ptc{\nu}} \beta^\ptc{\nu}  ) \\
   &= -\int d \Sigma_{\pt \ptc{\mu}} 
   \left( \beta^\ptc{\nu} \Theta^{\ptc{\mu}}_{~\ptc{\nu}}
   + \frac{1}{2} \Sigma^{\ptc{\mu}~\ptc{\rho}}_{~\ptc{\nu}} D_\ptc{\rho} \beta^\ptc{\nu} 
   + \nu J^\ptc{\mu} \right)
   + \frac{1}{2} \int d S_{\pt\ptc{\mu} \ptc{\rho}} \Sigma^{\ptc{\mu}~\ptc{\rho}}_{~\ptc{\nu}} \beta^\ptc{\nu}, 
    \label{eq:Surface}
  \end{split}
\end{equation}
where $dS_{\pt \ptc{\rho}\ptc{\mu}}$ denotes the surface element
for the $(d-1)$-dimensional spatial region $\Sigma_\pt$, 
and we used the Stokes' theorem (See e.g. Ref.~\cite{Poisson}),
\begin{equation}
  \int_\Sigma d\Sigma_{\pt \ptc{\mu}} D_\ptc{\rho} B^{\ptc{\rho}\ptc{\mu}} 
  = \int_{\partial \Sigma} dS_{\pt \ptc{\mu}\ptc{\rho}}
  B^{\ptc{\rho}\ptc{\mu}}, 
\end{equation}
satisfied for anti-symmetric tensors $B^{\ptc{\mu}\ptc{\nu}} = - B^{\ptc{\nu}\ptc{\mu}}$ 
to obtain the last line in Eq.~\eqref{eq:Surface}. 
If the fields fall off sufficiently rapidly as $|\bm{\px}|\to \infty$, 
we can neglect the surface term.
The first term in the last line in Eq.~\eqref{eq:Surface} reproduces 
the spatial part of the Euclidean action, and it seems
the second term is not necessary in our discussion.
However, we show that second term is rather important
since it gives us the correct spin connection in the imaginary-time direction.

In order to show that we obtain the proper spin connection
in thermal spacetime, let us first introduce the (inverse)
thermal vielbein $\tilde{e}_\ptc{\mu}^{~a} \ ( \tilde{e}_a^{~\ptc{\mu}})$ as
\begin{equation}
 \begin{split}
   \tilde{e}_\ptc{0}^{~a} &= e^\sigma u^a, \quad ~~
  \tilde{e}_\ptc{i}^{~a} = e_\ptc{i}^{~a} , \\
  \tilde{e}_a^{~\ptc{0}} &= e_a^{~\ptc{0}} \frac{e^{-\sigma}}{u^\ptc{0}}, \quad 
  \tilde{e}_a^{~\ptc{i}} = e_a^{~\ptc{i}} -
  e_a^{~\ptc{0}}\frac{u^\ptc{i}}{u^\ptc{0}}, \label{eq:tildee}
  \end{split}
\end{equation}
where the thermal vielbein satisfies relations 
\begin{equation}
 \tilde{g}_{\ptc{\mu}\ptc{\nu}}
  = \tilde{e}_\ptc{\mu}^{~a}\tilde{e}_\ptc{\nu}^{~b} \eta_{ab}, \quad 
  \eta^{ab} = \tilde{e}_\ptc{\mu}^{~a} \tilde{e}_\ptc{\nu}^{~b}
  \tilde{g}^{\ptc{\mu}\ptc{\nu}}, 
\end{equation}
and the inverse vielbein satisfies
$\delta_{\ptc{\mu}}^\ptc{\nu}
= \tilde{e}_{\ptc{\mu}}^{~a}\tilde{e}^{~\ptc{\nu}}_{a}$, 
$\delta_{a}^b= \tilde{e}^{~\ptc{\mu}}_a \tilde{e}_{\ptc{\mu}}^{~b}$.
Compared with the relations such as Eq.~\eqref{eq:MetricVielbein}
which the original vielbein satisfies, 
it is properly considered as the vielbein
associated with the emergent thermal spacetime. 
We also introduced the covariant derivative in thermal spacetime as
\begin{equation}
 \tilde{D}_\ptc{\mu}
  \equiv \tilde{\partial}_\ptc{\mu}
  -  i ( \tilde{\mathcal{A}}_\ptc{\mu} + \tilde{A}_\ptc{\mu})
  \quad \mathrm{with} \quad
   \tilde{\mathcal{A}}_\ptc{\mu}
  \equiv \frac{1}{2}  \tilde{\omega}_{\ptc{\mu}}^{~ab} \varSigma_{ab}.
  \label{eq:CovDerFermion2}
\end{equation}
with $\tilde{A}_\ptc{\mu} =( e^\sigma \mu, A_\ptc{i})$. 
Here we introduced the thermal spin connection
$\tilde{\omega}_\mu^{~ab}$ given by
\begin{align}
  \tilde{\omega}_\mu^{~ab} 
  &=\frac{1}{2} \tilde{e}^{a\nu}\tilde{e}^{b\rho} 
  (\tilde{C}_{\nu\rho\mu} - \tilde{C}_{\rho\nu\mu} - \tilde{C}_{\mu\nu\rho}), \\
  \tilde{C}_{\mu\nu\rho}
  &\equiv \tilde{e}_{\mu}^{~c}
  ( \tilde{\partial}_{\nu} \tilde{e}_{\rho c}
 - \tilde{\partial}_{\rho}\tilde{e}_{\nu c} ),
 \label{eq:ThermalRicci}
\end{align}
where we defined the thermal Ricci rotation coefficients
$\tilde{C}_{\mu\nu\rho}$. 
Here we put the tetrad postulate for the thermal spacetime:
\begin{equation}
 \tilde{D}_\mu e_\nu^{~a}
  = \tilde{\partial}_\mu \tilde{e}_\nu^{~a}
  - \tilde{\Gamma}^\rho_{~\mu\nu} \tilde{e}_\rho^{~a}
  + \tilde{\omega}_{\mu~b}^{~a} \tilde{e}_\nu^{~b} = 0,
  \label{eq:ThermalTetrad}
\end{equation}
where we used the Christoffel symbol $\tilde{\Gamma}^\rho_{~\mu\nu}$ composed of the thermal metric. 

Based on this setup, we demonstrate that the Masseiu-Planck functional is
again expressed in terms of the proper Euclidean action in the
emergent curved spacetime. 
First, we show that the thermal spin connection
for the imaginary-time direction 
is originated from the second term in the Eq.~\eqref{eq:Surface}. 
From Eq.~\eqref{eq:ThermalTetrad} we have 
\begin{equation}
  \tilde{\omega}_{\ptc{\mu}\ptc{\nu}\ptc{\rho}}
  \equiv -
  \tilde{e}_{\ptc{\nu}a}\tilde{\partial}_{\ptc{\mu}}\tilde{e}_{\ptc{\rho}}^{~a}
  +\tilde{\Gamma}_{\ptc{\nu}\ptc{\mu}\ptc{\rho}}
\end{equation}
with 
\begin{equation}
  \tilde{\Gamma}_{\ptc{\nu}\ptc{\mu}\ptc{\rho}}
   =\frac{1}{2} (\tilde{\partial}_{\ptc{\mu}} \tilde{g}_{\ptc{\nu}\ptc{\rho}}
   + \tilde{\partial}_{\ptc{\rho}}\tilde{g}_{\ptc{\mu}\ptc{\nu}}
   -\tilde{\partial}_{\ptc{\nu}}\tilde{g}_{\ptc{\mu}\ptc{\rho}} ).
\end{equation}
Paying attention to the fact that all of our parameters do not
depend on the imaginary time, and thus,
thermal metric and vielbein do not too,
we can express $\tilde{\omega}_{\ptc{0}\ptc{i}\ptc{j}}$ in terms of
the fluid vector $\beta^\mu(x)$ as 
\begin{equation}
  \tilde{\omega}_{\ptc{0}\ptc{i}\ptc{j}} 
  = \tilde{\Gamma}_{\ptc{i}\ptc{0}\ptc{j}}
  = - \frac{1}{2\beta_0}
  ( \partial_{\ptc{i}}\beta_{\ptc{j}} - \partial_{\ptc{j}}\beta_{\ptc{i}} )
\end{equation}
where we used Eq.~\eqref{eq:thermalMetric}. 
Together with $ ee^{~\ptc{0}}_a  =  \tilde{e}  \tilde{e}^{~\ptc{0}}_a$,
we obtain 
\begin{equation}
 \begin{split}
 &\quad \int_0^{\beta_0} d\tau 
   d^{d-1}\px \, e
  \left[ \frac{-i}{2} \ptc{\psi}
  \left( {{e}_a^{~\ptc{0}}}\gamma^a \ra{\partial_\tau}
  - \la{\partial_\tau} e_a^{~\ptc{0}}\gamma^a \right) \psi 
  - \beta_0^{-1} \left(
  \frac{1}{2} \Sigma^{\ptc{0}~\ptc{\rho}}_{~\ptc{\nu}}
  D_\ptc{\rho} \beta^{\ptc{\nu}}
  +  \nu J^\ptc{0}
  \right)
  \right]\\
  &  = \int_0^{\beta_0}  d^d \tilde{x} \, \tilde{e}
  \left[
  -\frac{1}{2} \ptc{\psi}
  \left( 
  \tilde{e}_a^{~\ptc{0}} \gamma^{a} i\ra{\partial_\tau}
  - i\la{\partial_\tau} \tilde{e}_a^{~\ptc{0}}\gamma^a \right) \psi
  + \frac{i}{4}
  \tilde{e}_a^{~\ptc{0}}
  \ptc{\psi} \{\gamma^a,\varSigma_{bc} \} \psi 
  \tilde{\omega}_{\ptc{0}}^{~bc}
  + i \tilde{e}_a^{~0}e^\sigma \mu \ptc{\psi}\gamma^a \ptc{\psi}
  \right] \\
  & = \int_0^{\beta_0}  d^d \tilde{x} \, \tilde{e}
  \left[
  -\frac{1}{2} \ptc{\psi}
  \left\{
  \tilde{e}_a^{~\ptc{0}} \gamma^{a}
  \Big( \ra{\tilde{\partial}_\ptc{0}}
  - i (\tilde{\mathcal{A}}_\ptc{0} + \tilde{A}_\ptc{0}) \Big) 
  - \Big( \la{\tilde{\partial}_\ptc{0}}
  + i (\tilde{\mathcal{A}}_\ptc{0} + \tilde{A}_\ptc{0})
  \Big)
  \tilde{e}_a^{~\ptc{0}}\gamma^a \right\} \psi
  \right] \\
  & = \int_0^{\beta_0}  d^d \tilde{x} \, \tilde{e}
  \left[
  -\frac{1}{2} \ptc{\psi}
  \left( 
  \tilde{e}_a^{~\ptc{0}} \gamma^{a}
 \ra{\tilde{D}}_\ptc{0}
  - \la{\tilde{D}}_\ptc{0}
   \tilde{e}_a^{~\ptc{0}}\gamma^a \right) \psi
  \right] ,
 \end{split}
\end{equation}
where we used the fact that spin angular momentum tensor
$\Sigma^{\mu\nu\rho}$ defined in Eq.~\eqref{eq:SpinAM}
is completely antisymmetric with respect to its indices. 
The last line shows that we obtain the proper
covariant imaginary-time derivative
with the torsion-free thermal spin connection. 
Furthermore, we can prove that the spatial components of the thermal
spin connection are in accordance with the original ones:
\begin{equation}
  \bar{\psi}\{\gamma_a,\varSigma_{bc} \} \psi
  \tilde{e}^{a\ptc{i}}\omega_{\ptc{i}}^{~bc}
  = \bar{\psi}\{\gamma_{a},\varSigma_{bc} \} \psi
  \tilde{e}^{a\ptc{i}}\tilde{\omega}_{\ptc{i}}^{~bc}.
  \label{eq:proof}
\end{equation}
In order to prove this identity we need to evaluate
\begin{equation}
  \tilde{\omega}_{\ptc{\mu}\ptc{\nu}\ptc{\rho}}
  =\frac{1}{2}(\tilde{C}_{\ptc{\nu}\ptc{\rho}\ptc{\mu}}-\tilde{C}_{\ptc{\rho}\ptc{\nu}\ptc{\mu}}-\tilde{C}_{\ptc{\mu}\ptc{\nu}\ptc{\rho}}),
\end{equation}
with the thermal Ricci rotation coefficients $\tilde{C}_{\mu\nu\rho}$
defined in Eq.~\eqref{eq:ThermalRicci}. 
Since the spatial partial derivative and the thermal vielbein is
unchanged in thermal spacetime:
$\tilde{\partial}_\ptc{i}={\partial}_\ptc{i}$ and  
$\tilde{e}_{\ptc{i}}^{~a}={e}_{\ptc{i}}^{~a}$,
we find that some spatial components of the thermal spin connection
are identical to the original ones: 
$\tilde{\omega}_{\ptc{i}\ptc{j}\ptc{k}}={\omega}_{\ptc{i}\ptc{j}\ptc{k}}$.
Then, in order to compare $ \tilde{\omega}_\ptc{i}^{~bc}$
with ${\omega}_\ptc{i}^{~bc}$,
we express the original spin connection ${\omega}_\ptc{i}^{~bc}$
in terms of the thermal vielbein $\tilde{e}_{\ptc{\mu}}^{~a}$.
For that purpose, we decompose $ {\omega}_\ptc{i}^{~bc}$
into their components as
\begin{equation}
 \omega_\ptc{i}^{~bc}
  = \omega_{\ptc{i}\ptc{\mu}\ptc{\nu}} e^{b\ptc{\mu}} e^{c\ptc{\nu}}
  = ( e^{b\ptc{0}}{e}^{c\ptc{j}} - e^{b\ptc{j}}{e}^{c\ptc{0}})
  \omega_{\ptc{i}\ptc{0}\ptc{j}}
  + e^{b\ptc{j}}{e}^{c\ptc{k}} \omega_{\ptc{i}\ptc{j}\ptc{k}}.
\end{equation}
Using Eq.~\eqref{eq:tildee},
we can express the inverse vielbein in terms of
the inverse thermal vielbein as $ e^{a\ptc{i}}=
\tilde{e}^{a\ptc{i}}+e^{a\ptc{0}}{u^\ptc{i}}/{u^\ptc{0}}$. 
Therefore, the second term becomes
\begin{equation}
 \begin{split}
  e^{b\ptc{j}}{e}^{c\ptc{k}} \omega_{\ptc{i}\ptc{j}\ptc{k}}
  &=  \left( \tilde{e}^{b\ptc{j}}
  + e^{b\ptc{0}}\frac{u^\ptc{j}}{u^\ptc{0}} \right)
  \left( \tilde{e}^{c\ptc{k}}
  + e^{c\ptc{0}}\frac{u^\ptc{k}}{u^\ptc{0}} \right)
  \omega_{\ptc{i}\ptc{j}\ptc{k}}\\
  &= ( e^{b\ptc{0}}\tilde{e}^{c\ptc{j}}
  - \tilde{e}^{b\ptc{j}}{e}^{c\ptc{0}})
  \frac{u^\ptc{k}}{u^{\ptc{0}}}\omega_{\ptc{i}\ptc{k}\ptc{j}}
   + \tilde{e}^{b\ptc{j}} \tilde{e}^{c\ptc{k}} \omega_{\ptc{i}\ptc{j}\ptc{k}},
 \end{split}
\end{equation}
where we used the antisymmetric property of the spin connection:
$\omega_{\ptc{i}\ptc{j}\ptc{k}} = -\omega_{\ptc{i}\ptc{k}\ptc{j}}$. 
Furthermore, noting
${e}^{b\ptc{0}}{e}^{c\ptc{j}} -{e}^{b\ptc{j}}{e}^{c\ptc{0}}
={e}^{b\ptc{0}}\tilde{e}^{c\ptc{j}} - \tilde{e}^{b\ptc{j}}{e}^{c\ptc{0}}$,
we obtain
\begin{equation}
 \begin{split}
  \omega_\ptc{i}^{~bc}
  &=
  ( e^{b\ptc{0}}\tilde{e}^{c\ptc{j}} - \tilde{e}^{b\ptc{j}} e^{c\ptc{0}})
  \frac{u^\ptc{\mu}}{u^{\ptc{0}}}\omega_{\ptc{i}\ptc{\mu}\ptc{j}}
  + \tilde{e}^{b\ptc{j}} \tilde{e}^{c\ptc{k}} \omega_{\ptc{i}\ptc{j}\ptc{k}} \\
  &= ( \tilde{e}^{b\ptc{0}}\tilde{e}^{c\ptc{j}}
  - \tilde{e}^{b\ptc{j}}\tilde{e}^{c\ptc{0}})
  { e^{\sigma}u^\ptc{\mu} }\omega_{\ptc{i}\ptc{\mu}\ptc{j}}
  + \tilde{e}^{b\ptc{j}} \tilde{e}^{c\ptc{k}}
  \tilde{\omega}_{\ptc{i}\ptc{j}\ptc{k}}.
 \end{split}
\end{equation}
In contrast, $\tilde{\omega}_\ptc{i}^{~bc}$ is simply expressed as
\begin{equation}
  \tilde{\omega}_\ptc{i}^{~bc}
  =  (\tilde{e}^{b\ptc{0}}\tilde{e}^{c\ptc{j}}
  - \tilde{e}^{b\ptc{j}}\tilde{e}^{c\ptc{0}})
  \tilde{\omega}_{\ptc{i}\ptc{0}\ptc{j}}
  + \tilde{e}^{b\ptc{j}}\tilde{e}^{c\ptc{k}} \omega_{\ptc{i}\ptc{j}\ptc{k}}.
\end{equation}
Therefore, the difference between them is given by 
\begin{equation}
 \begin{split}
  \bar{\psi}\{\gamma_a,\varSigma_{bc} \} \psi
  \tilde{e}^{a\ptc{i}}( {\omega}_\ptc{i}^{~bc} -\tilde{\omega}_\ptc{i}^{~bc})
  &= 2\bar{\psi}\{\gamma_a,\varSigma_{bc} \} \psi
  \tilde{e}^{a\ptc{i}}\tilde{e}^{b\ptc{0}}\tilde{e}^{c\ptc{j}} 
  (e^{\sigma}u^\ptc{\mu} \omega_{\ptc{i}\ptc{\mu}\ptc{j}}-
  \tilde{\omega}_{\ptc{i}\ptc{0}\ptc{j}})\\
  &= \bar{\psi}\{\gamma_a,\varSigma_{bc} \} \psi
  \tilde{e}^{a\ptc{i}}\tilde{e}^{b\ptc{0}}\tilde{e}^{c\ptc{j}} 
  (e^{\sigma}u^\ptc{\mu} {C}_{\ptc{\mu}\ptc{j}\ptc{i}}
  - \tilde{C}_{\ptc{0}\ptc{j}\ptc{i}}),
  \label{eq:proof2}
 \end{split}
\end{equation}
where we dropped the symmetric part of $\omega_{\ptc{i}\ptc{\mu}\ptc{j}}$
under $\ptc{i}\leftrightarrow\ptc{j}$ in the second line
because $\ptc{\psi} \{\gamma_a, \varSigma_{bc}\}$ is completely
antisymmetric for its indices. 
By the way,
noting that the imaginary-time derivative of the thermal vielbein vanishes, 
the definiton of $\tilde{C}_{\ptc{\mu}\ptc{\nu}\ptc{\rho}}$ 
in Eq.~\eqref{eq:ThermalRicci} leads to 
\begin{equation} 
  \tilde{C}_{\ptc{0}\ptc{i}\ptc{j}}
  = \tilde{e}_{\ptc{0}}^{~c}
  ({\partial}_{\ptc{i}}{e}_{\ptc{j} c}-{\partial}_{\ptc{j}}{e}_{\ptc{i} c})
  = e^{\sigma}u^{\ptc{\mu}}C_{\ptc{\mu}\ptc{i}\ptc{j}}. 
  \label{eq:C0ij}
\end{equation}
Therefore, we find Eq.~\eqref{eq:proof2} vanishes, 
and, thus, Eq.~\eqref{eq:proof} is proved.

In conclusion, we finally get the following expression for the
Masseiu-Planck functional: 
\begin{equation}
  \Psi[\pt;\lambda,A_\ptc{\mu}] = \log
  \int \mathcal{D}\psi\mathcal{D}\bar{\psi}\,
  e^{S[\psi,\ptc{\psi}; \lambda,A_\ptc{\mu}]}, 
\end{equation}
with 
\begin{equation}
  \begin{split}
    S[\psi,\ptc{\psi};\lambda,A_\ptc{\mu}]
   &= \int_0^{\beta_0} d\tau \int
   d^{d-1}\ptc{x}\tilde{e}\left[-\frac{1}{2} \ptc{\psi}(\tilde{e}_{m}^{~\ptc{\mu}}\gamma^m\ra{\tilde{D}}_{\ptc{\mu}}
   -\la{\tilde{D}}_{\ptc{\mu}}\tilde{e}_{m}^{~\ptc{\mu}}\gamma^m)\psi 
   - m \ptc{\psi}\psi
   \right]   \\
   &= \int _0^{\beta_0} d^d \tilde{x} \,
   \tilde{e} \Lcal(\psi,\ptc{\psi},
   \tilde{\partial}_\rho \psi, \tilde{\partial}_\rho \ptc{\psi};
   \tilde{e}_\ptc{\mu}^{~a},\tilde{A}_\ptc{\mu})   .  
  \end{split}
\end{equation}
This shows that the Masseiu-Planck functional is again written in terms of the Euclidean action 
in the emergent curved spacetime in the same way as fields with the integer spin. 
Although it is expressed by the thermal vielbein, there exists no torsion, 
and the structure of the emergent thermal space is completely same as the previous case. 
It should be emphasized that the resulting action contains the
proper covariant derivative in thermal spacetime, so that it formally possesses full diffeomorphism
invariance in the emergent thermal spacetime, which is discussed in the
next section.
Note that, as is the case for the thermal metric,
the thermal vielbein $\tilde{e}_\ptc{\mu}^{~a}$ and original
vielbein $e_\ptc{\mu}^{~a}$ coincide with each other in the hydrostatic gauge:
$\tilde{e}_\ptc{\mu}^{~a} = e_\ptc{\mu}^{~a}\big|_\mathrm{hs}$. 
In fact, because of the hydrostatic gauge condition
$e^\sigma u^\ptc{\mu} = \delta^\ptc{\mu}_\ptc{0}$,
we can immediately show the non-trivial part of this relation as
$\tilde{e}_\ptc{0}^{~a} = e^\sigma u^\ptc{\mu}
e_\ptc{\mu}^{~a} = e_\ptc{0}^{~a}\big|_\mathrm{hs}$ from Eq.~\eqref{eq:tildee}. 

\section{Symmetries of emergent thermal spacetime}
\label{sec:Symmetry}
In the previous section, we have shown that the Masseiu-Planck
functional for any quantum field with the spin $0$, $1/2$, and $1$ 
is written in the language of the path integral
\eqref{eq:PathIntegralFormula} in the thermally emergent
curved spacetime background, 
whose line element again has the form of the ADM metric:
\begin{equation}
  d\tilde{s}^2 = -(\tilde{N}d\tilde{t})^2 +\gamma_{\ptc{i}\ptc{j}}(\tilde{N}^\ptc{i}d\tilde{t}+d\px^\ptc{i})(\tilde{N}^\ptc{j}d\tilde{t}+d\px^\ptc{j}),
  \label{eq:ThermalSpacetime}
\end{equation}
with the thermal lapse function $\tilde{N}$ and thermal shift vector
$\tilde{N}^{\ptc{i}}$ defined in Eq.~\eqref{eq:tildeLapseShift}.
We have also considered the conserved charge current
which couples to the external $U(1)$ gauge field. 
It is described by the presence of a background $U(1)$ gauge connection 
which is slightly modified by the local chemical potential $\mu(x)$ as 
\begin{equation}
 \tilde{A} = \tilde{A}_\ptc{0} d\tilde{t} + \tilde{A}_\ptc{i} d\px^\ptc{i},
  \label{eq:GaugeConnection}
\end{equation}
with the background gauge field $\tilde{A}_\ptc{\mu}$
defined in Eq.~\eqref{eq:tildeA}.
Since the time component of the original external field does not
appear in our construction, $\tilde{A}_\ptc{0}$ only contains
the local chemical potential. 
Therefore, the structure of the emergent thermal spacetime
and gauge connection is completely determined
by the hydrodynamic configurations $\lambda^a (x)$.

As can been seen from Eq.~\eqref{eq:PathIntegralFormula},
we have formally have full diffeomorphsim invariance and 
$U(1)$ gauge invariance in the thermal spacetime 
since the action contains the proper covariant derivative%
\footnote{
 Of course, we should understand symmetry transformations, 
 keeping in mind that one coordinate is imaginary.
 }.
However, we have to pay attention to the fact
that all of the thermal metric, vielbein, and external gauge field 
are not dependent on the imaginary time.
One way to treat this is simply to perform path integrals 
as if we have full invariance, and neglect the imaginary-time
dependence at the end. 
For example, we can construct the derivative expansion of $\Psi[\pt;\lambda]$
with the use of invariants such as loop like objects:
$\displaystyle{\oint d\tilde{s},~
\oint \tilde{A}_\mu d\tilde{x}^\mu}$,
and derivatives of them such as the Ricci scalar for the
$d$-dimensional thermal spacetime: $\tilde{R}$, 
as building blocks.

However, the symmetry properties can be expressed in a bit different manner. 
In this section, taking another way,
we elucidate the symmetries of this thermally
emergent curved spacetime, and the background $U(1)$ gauge connection:
Kaluza-Klein gauge symmetry, spatial difeomorphism symmetry,
and gauge symmetry. 
First, we show the most prominent symmetry property related to the our imaginary-time formalism, 
that is, the Kaluza-Klein gauge symmetry of the Masseiu-Planck functional in Sec.~\ref{sec:KaluzaKlein}. 
We next see that it also has ($d-1$)-dimensional spatial diffeomorphism invariance in Sec.~\ref{sec:SpatialDiffeo}. 
In addition to these spacetime symmetries, we see the symmetric properties 
for the background $U(1)$ gauge connection in Sec.~\ref{sec:Gauge}.
These symmetry arguments lay out a foundation to derive
the transport properties of locally thermalized matters,
and thus, hydrodynamic equations as discussed in
Ref.~\cite{Hayata:2015lga} (See also
Refs.~\cite{Banerjee:2012iz,Jensen:2012jh,Haehl:2015pja,Crossley:2015evo}).

\subsection{Kaluza-Klein gauge symmetry}
\label{sec:KaluzaKlein}
First of all, we point out that the structure of the emergent thermal spacetime is invariant under the global imaginary-time translation, 
since the thermodynamic parameters $\lambda^a(x)$ such as the local temperature and fluid-four velocity 
do not depend on the imaginary time $\tau$, and thus $\tilde{t}=-i\tau$.
Furthermore, we also have a local symmetry by the spatial coordinate-dependent redefinition of the imaginary time. 
In order to demonstrate this symmetry, we rewrite the line element
in thermal spacetime from the ADM form to the Kaluza-Klein form as
\begin{equation}
  d\tilde{s}^2 
   = -e^{2\sigma}(d\tilde{t}+a_{\ptc{i}}d\px^{\ptc{i}})^2
   +\gamma'_{\ptc{i}\ptc{j}}d\px^\ptc{i} d\px^\ptc{j} , 
   \label{eq:KKGeometry}
\end{equation}
where we defined $a_{\ptc{i}}\equiv  -e^{-\sigma}u_\ptc{i} $, ${\gamma'}_{\ptc{i}\ptc{j}}\equiv\gamma_{\ptc{i}\ptc{j}}+u_\ptc{i}u_\ptc{j}$, and
we used $\tilde{g}_{\ptc{0}\ptc{0}} = -\tilde{N}^2 +\tilde{N}_\ptc{i}\tilde{N}^\ptc{i} =-e^{2\sigma}$.
In this parametrization, the square root of determinant of the thermal metric becomes 
$\sqrt{-\tilde{g}}= \tilde{N}\sqrt{\gamma}=e^{\sigma}\sqrt{\gamma'}$.
This parametrization of the line element was discussed
in the hydrostatic generating functional method~\cite{Banerjee:2012iz},
and we can discuss the symmetry properties of the Massieu-Planck functional
in a similar manner. 
Following Ref.~\cite{Banerjee:2012iz},
we can easily see that this line element is invariant
under the local transformation,
or the Kaluza-Klein gauge transformation:
\begin{equation}
 \begin{cases}
  \tilde{t} \to \tilde{t} + \chi(\bm{\px}), \\
  \bm{\px} \to \bm{\px}, \\
  a_\ptc{i}(\bm{\px}) \to a_\ptc{i}(\bm{\px}) - \partial_\ptc{i} \chi(\bm{\px}), 
  \label{eq:KKTrans}
 \end{cases}
\end{equation}
where $\chi(\bm{\px})$ is an arbitrary function of the spatial coordinates.
We note that the original induced metric $\gamma_{\ptc{i}\ptc{j}}$ nonlinearly transforms under this transformation 
since $\gamma'_{\ptc{i}\ptc{j}}$ does not change, so that $\gamma$ is not Kaluza-Klein gauge invariant.

This symmetry enables us to restrict possible terms that appear in the
construction of the Massieu-Planck functional in the same way
as the hydrostatic generating functional method \cite{Banerjee:2012iz}.
In fact, while this symmetry does not restrict a dependence
on the dilaton sector, which is the local temperature $e^{\sigma(x)}=\beta(x)/\beta_0$, 
it strongly does on the thermal Kaluza-Klein gauge field $a_\ptc{i}$. 
For example, $a_\ptc{i}$ appears in the Massieu-Planck functional only through the gauge invariant combination such as
the field strength $f_{\ptc{i}\ptc{j}}$ defined by  
\begin{equation}
  f_{\ptc{i}\ptc{j}} \equiv \partial_\ptc{i} a_\ptc{j} - \partial_\ptc{j} a_\ptc{i}.
  \label{eq:KKstrength}
\end{equation}
As is shown in Sec.~\ref{sec:Gauge}, the Kaluza-Klein symmetry also affects
how the Masseiu-Planck functional depends on the external gauge field $A_{\ptc{i}}$.

\subsection{Spatial diffeomorphism symmetry}
\label{sec:SpatialDiffeo}
As is developed in Sec.~\ref{sec:ADM}, utilizing the ADM decomposition,
we introduced the spatial-coordinate system $\bm{\px}=\bm{\px}(x)$ on a spacelike hypersurface $\Sigma_\pt$. 
The spatial coordinate system is described by the original induced metric $\gamma_{\ptc{i}\ptc{j}}$, 
or equivalently the modified one $\gamma_{\ptc{i}\ptc{j}}'$.

If we recall the simple fact that Physics does not depend on our choice of the spatial-coordinate system 
$\bm{\px} = \bm{\px}(x)$, we can immediately see that
the Masseiu-Planck functional $\Psi[\pt,\lambda]$ is 
invariant under the $(d-1)$-dimensional spatial diffeomorphism
\begin{equation}
 \bm{\px} \rightarrow \bm{\px}'(\bm{\px}).
  \label{eq:SpatialDiffeo}
\end{equation}
This spatial diffeomorphism invariance also restricts possible terms that could appear in 
the construction of the Massieu-Planck functional. 
For example, $\gamma'$ appears only in combination with $d^{d-1}\px$, i.e., $d^{d-1} \px \sqrt{\gamma'}=d\Sigma_\pt N e^{-\sigma}$. 
Note that we use $\sqrt{\gamma'}$ instead of $\sqrt{\gamma}$. 
This is because the modified $\gamma'$ is Kaluza-Klein gauge invariant
while the original one $\gamma$ is not.

\subsection{Gauge connection and gauge symmetry}
\label{sec:Gauge}
In the presence of the conserved $U(1)$ current coupled to the external field $A_\ptc{i}$, 
we have also the background $U(1)$ gauge connection \eqref{eq:GaugeConnection} 
at the same time as the emergent thermal spacetime \eqref{eq:ThermalSpacetime}, or \eqref{eq:KKGeometry}. 
As is already mentioned,
we do not have the time-component of the original external field $A_\ptc{0}$, 
and the Masseiu-Planck functional is invariant under 
\begin{equation}
  A_\ptc{i} (\bm{\px}) \to A_\ptc{i}(\bm{\px}) + \partial_\ptc{i} \alpha (\bm{\px}).
  \label{eq:GaugeTr}
\end{equation}

While the local chemical potential $e^\sigma \mu$ is Kaluza-Klein gauge invariant, 
$A_\ptc{i}$ is not from the same reason that 
the original induced metric $\gamma_{\ptc{i}\ptc{j}}$ is not. 
It is, then, convenient to rewrite the gauge connection \eqref{eq:GaugeConnection}
in the similar way to Eq.~\eqref{eq:KKGeometry} 
as follows:
\begin{equation}
  \tilde{A}= \tilde{A}_\ptc{0}' (d\tilde{t} + a_\ptc{i} d\px^\ptc{i}) 
  + \tilde{A}_\ptc{i}' d\px^\ptc{i}, 
  \label{eq:GaugeConnection2}
\end{equation}
where we defined the modified gauge field $\tilde{A}'_\ptc{\mu}$
in thermal spacetime as
\begin{equation}
  \begin{split}
   \tilde{A}_\ptc{0}' &\equiv \tilde{A}_\ptc{0} = e^\sigma \mu, \\
   \tilde{A}_\ptc{i}'
   &\equiv \tilde{A}_\ptc{i} - \tilde{A}_\ptc{0} a_\ptc{i}
   = A_\ptc{i} - e^\sigma \mu a_\ptc{i}.  
  \end{split}
\end{equation}
From Eq.~\eqref{eq:GaugeConnection2},
it becomes clear that this modified gauge field $\tilde{A}_\ptc{\mu}'$
remains invariant 
under the Kaluza-Klein gauge transformation \eqref{eq:KKTrans}, since the combination $d\tilde{t}+ a_\ptc{i}d\px^\ptc{i}$ 
is unchanged.
Moreover, this modified background gauge field behaves 
in the same manner as the original one under the gauge transformation in Eq.~\eqref{eq:GaugeTr},
We, therefore, rephrase that the Masseiu-Planck functional is invariant under  
\begin{equation}
 \tilde{A}_\ptc{i}' (\bm{\px})
  \to \tilde{A}_\ptc{i}'(\bm{\px}) + \partial_\ptc{i} \alpha (\bm{\px}).
\end{equation}
From this useful property, we should use the
modified gauge field $\tilde{A}_\ptc{\mu}'$
instead of the original one $\tilde{A}_\ptc{\mu}$ in thermal spacetime.

\section{Summary and Outlook}
\label{sec:Summary}
\begin{figure}[t]
 \centering
 \includegraphics[width=0.9\linewidth]{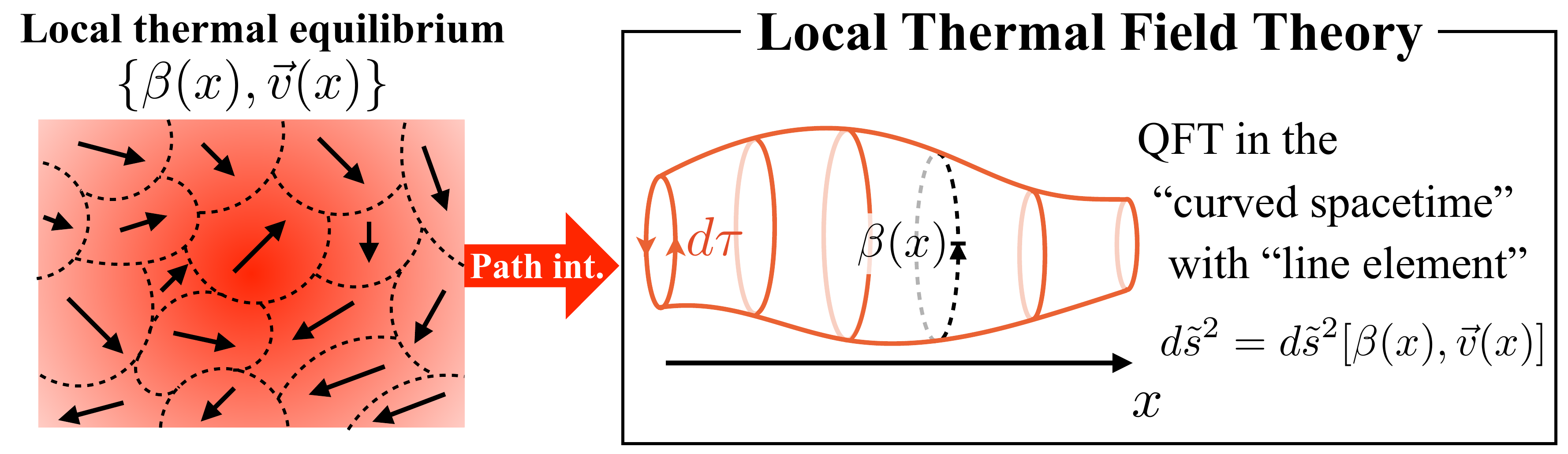}
 \caption{
 Schematic picture of the imaginary-time formalism for local thermal equilibrium. 
} 
\label{Fig:local-TQFT}
\end{figure}

In this paper, the imaginary-time formalism for systems
under local thermal equilibrium, 
in which the density operator has a form of the local Gibbs distribution, 
has been presented on the basis of the path-integral formulation.  
After a meticulous preparation in Sec.~\ref{sec:LocalGibbs}, 
we have shown that the Masseiu-Planck functional $\Psi[\pt;\lambda]$ 
plays a role as the generating functional of the expectation values of 
the conserved current operators over the local Gibbs distribution
in Sec.~\ref{sec:MasseiuPlanck}. 
Indeed, we have derived the variational formula
given in Eq.~\eqref{eq:VariationFormula1}
without choosing any particular coordinate system, 
and given in Eq.~\eqref{eq:VariationFormula2} with the hydrostatic gauge 
in which the hydrodynamic configurations looks like entirely at rest
due to the hydrostatic gauge fixing condition~\eqref{eq:HydroStaticCond}. 
Furthermore,
through the detailed analysis on representative examples of relativistic
quantum fields such as the scalar fields, Dirac field, and gauge fields
in Sec.~\ref{sec:PathIntegral}, 
we have reached the conclusion that the Masseiu-Planck functional
$\Psi[\pt;\lambda]$ is written in terms of the path integral
of the Euclidean action in the thermally emergent curved spacetime
as is given in Eq.~\eqref{eq:PathIntegralFormula}. 
Our result is schematically summarized in Fig.~\ref{Fig:local-TQFT} 
(Compare with Fig.~\ref{Fig:imaginary-time}).
This emergent curved spacetime with 
one imaginary-time direction, and $(d-1)$-spatial directions  
is described by the use of the thermal metric \eqref{eq:thermalMetric},
thermal vielbein \eqref{eq:tildee}, and external gauge field \eqref{eq:tildeA}, 
which are completely determined
by configurations of the local thermodynamic
variables $\lambda^a(x)$ on hypersurfaces. 
As is discussed in Sec.~\ref{sec:Symmetry}, 
our action is formally equipped with
full diffeomorphism invariance and full gauge invariance
in thermal spacetime with the imaginary-time independent metric,
vielbein, and external gauge field.  
This symmetry property eventually leads to
the fact that the Masseiu-Planck functional
possesses the notable intrinsic symmetries
associated with the local Gibbs distribution. 
They are the Kaluza-Klein gauge symmetry
under Eq.~\eqref{eq:KKTrans},
the spatial diffeomorphism symmetry~\eqref{eq:SpatialDiffeo}, 
and the gauge symmetry for the background gauge field~\eqref{eq:GaugeTr}. 
These results provide a general microscopic justification of 
the generating functional method to construct
nondissipative hydrodynamic constitutive relations 
based on the underlying quantum field theories.
In addition, it should be emphasized that our formulation is not
restricted to the hydrostatic configurations, in which
the fluid vector $\beta^\mu(x)$ becomes a killing vector
and the chemical potential gradient $\nabla_\mu \nu (x)$
balance an external electric field, 
but also applicable to any configuration of $\lambda^a(x)$%
\footnote{
We note that while Kaluza-Klein gauge invariance
in the hydrostatic partition function method~\cite{Banerjee:2012iz}
is originated from the hydrostatic condition for the real time direction,
our invariance is inherent in the local Gibbs distribution
with the imaginary time direction.
}. 
Therefore, our formulation provides a robust generalization of
the generating functional method,
and lays out a solid basis to discuss hydrodynamics
on the ground of microscopic quantum field theories.

There are several prospects on future research based on our approach. 
One is an explicit evaluation of the Masseiu-Planck functional based on 
quantum field theories. 
Although the symmetry arguments strongly restricts the possible form of the 
Masseiu-Planck functional, it does not completely determine its
functional dependence on thermodynamic parameters $\lambda^a (x)$,
in which information on thermodynamic and transport properties like the 
equation of state is contained. 
Since our formulation, in conjunction with the perturbative 
diagramatic approach, nonperturbative lattice simulations, or
holographic calculations, 
enables us to evaluate the derivative corrections to the Masseiu-Planck
functional, we can explicitly calculate the nondissipative
derivative corrections to relativistic hydrodynamics.
Although evaluating such a nondissipative correction has been discussed
in Refs.~\cite{Moore:2012tc,Becattini:2015nva}
in some restricted situations, our formulation
gives a general way to evaluate them, and shed light on
the thermodynamic and transport properties based on quantum field theories.
Also it is interesting to consider the case in which certain parts of 
derivatives of thermodynamic parameters such as a vorticity and 
temperature gradient does not take small values. 
Although we cannot perform naive perturbative expansions in this case, 
it may be possible to evaluate them in a nonperturbative manner.

Another interesting direction is a generalization to the
nonrelativistic quantum field theory and its application to
condensed matter physics. 
We are able to extend our approach to nonrelativistic theories, 
in which the structure of the emergent thermal spacetime may be given 
by the so-called Newton-Cartan geometry~\cite{Son:2013rqa,Geracie:2014nka,Gromov:2014vla,Jensen:2014aia,Jensen:2014ama,Jensen:2014wha}. 
We have shown that there does not exist the emergent 
thermal torsion in relativistic theories even if we consider spinor fields. 
However, in the nonrelativistic theory, we may reach at the redundancy of 
the description of nonrelativistic curved spacetime geometry and 
external gauge fields, which allows the emergent torsion to
appear.  
Although invariance coming from this redundancy has already been discussed in
Refs.~\cite{Jensen:2014aia,Jensen:2014ama,Jensen:2014wha}, it is
interesting to uncover the origin of the nonrelativistic geometry
based on our local Gibbs ensemble approach.
Generalization to the nonrelativistic systems and its application to
condensed matter physics are left for future works.

\acknowledgements
I am really grateful to Yoshimasa Hidaka for collaboration
in the early stage of this work and subsequent stimulating discussions.  
I also thank T. Hatsuda for useful comments and
careful reading of the whole manuscript,
F. Beccattini, T. Hayata, P. Kovtun, K. Mameda, T. Noumi, M. Rangamani,
S. Ryu, A. Shitade, Y. Tanizaki, N. Yamamoto, and Y. Yokokura for
useful discussions. 
M.H. was supported by the Special Postdoctoral Researchers Program
at RIKEN. This work was partially supported by the RIKEN iTHES Project.

\bibliography{partitionFunction}
\end{document}